\documentclass[a4paper,11pt]{article}
\usepackage{jcappub} 

\usepackage{graphicx}
\usepackage{amsmath, amssymb}
\usepackage{appendix}
\usepackage{verbatim, xcolor}
\usepackage{booktabs}

\title{\boldmath A Landscape of Cosmological Decoherence}

\author[a]{Bret Underwood,}
\author[b,c]{S. Shajidul Haque}

\affiliation[a]{Department of Physics,\\
Pacific Lutheran University,\\
Tacoma, WA 98447, United States}

\affiliation[b]{Department of Mathematics and Applied Mathematics,\\
University of Cape Town, South Africa}

\affiliation[c]{The National Institute for Theoretical and Computational Sciences, \\
Private Bag X1, Matieland, South Africa}

\emailAdd{bret.underwood@plu.edu}
\emailAdd{shajid.haque@uct.ac.za}

\abstract{
Current observations constrain primordial perturbations to be adiabatic, approximately Gaussian, and nearly-scale invariant. However, a generic mixed state satisfying these constraints has additional unconstrained degrees of freedom, which can be parameterized by the purity of the state and its momentum variance. This allowable parameter space reveals a unified geometric landscape of mixed states, allowing us to map and relate distinct models of decoherence and their respective pointer bases. Within this landscape we show that decoherence models that ``classicalize'' -- in the sense of admitting a regular, positive-definite Glauber-Sudarshan $P$-function -- must actively inject momentum into the system, exceeding that of the vacuum. The enhanced momentum sources the decaying mode of the curvature perturbation, backreacting on the Newtonian potential and radiation-era geometry. While this mode decays away fast enough to preserve the temporal coherence of the CMB acoustic peaks, requiring the potential to remain within linear perturbation theory places a model-independent bound on the momentum variance generated by any model of decoherence. This bound is definitively violated by decohered thermal states with more than $14$ e-folds of inflation, while a strong dependence on the number of e-folds restricts amplitude-basis decoherence to fewer than roughly $64$ e-folds of inflation in order to stay in the linear regime. Altogether, we present a unifying framework for evaluating the quantum-to-classical transition of the early universe.
}

\begin{document}

\maketitle
\flushbottom

\section{Introduction}
\label{sec:intro}

Cosmic inflation provides an elegant and robust mechanism for the origin of structure in the universe: macroscopic density perturbations arise from microscopic quantum fluctuations stretched to cosmic scales during an epoch of accelerated expansion \cite{Starobinsky:1980te,Guth:1980zm,Linde:1981mu,Albrecht:1982wi,Linde:1983gd}. 
This paradigm is spectacularly supported by high-precision observations of the Cosmic Microwave Background (CMB) and large-scale structure \cite{Planck:2018vyg,Planck:2018jri,Planck:2019kim}. 
In standard cosmological calculations, once these modes exit the Hubble horizon they are treated as classical stochastic variables whose probability distributions are governed by their quantum mechanical variances \cite{PhysRevD.50.4807,Polarski:1995jg, Lesgourgues:1996jc, Kiefer:1998qe, Kiefer:1998jk, Kiefer:1998pb, Kiefer:2006je, Kiefer:2008ku}.

However, the fundamental nature of these primordial perturbations -- whether they remain genuine pure quantum mechanical states, truly classical stochastic variables, or something in-between -- remains an open question (see \cite{Martin:2015qta, Martin:2019wta,Martin:2022kph,Micheli:2022tld} for some discussion).
The application of quantum information theory to the early universe has recently provided a powerful dictionary for addressing this. For instance, the rich dynamical evolution and highly squeezed structure of the pure and decohered inflationary state can be fruitfully characterized using measures like cosmological complexity \cite{Bhattacharyya:2020rpy,Bhattacharyya:2020kgu,Haque:2021kdm,Haque:2021hyw,Bhattacharyya:2024duw, Bhattacharyya:2025cxv,Bhattacharyya:2025lsc, Bhattacharyya:2026qef,Beetar:2023mfn,Bhattacharyya:2022rhm,Bhattacharyya:2021fii}.
Yet, to understand how this quantum structure eventually yields the classical universe we observe today, we must utilize the formalism of environmental decoherence. 
Decoherence occurs when the observable, long-wavelength modes of the primordial perturbations interact with an environment, such as short-wavelength modes, higher-order gravitational nonlinearities, or other spectator fields (see \cite{Burgess:2006jn,Burgess:2014eoa,Nelson:2016kjm,Shandera:2017qkg,Burgess:2022nwu,Burgess:2024eng,deKruijf:2024ufs,Lopez:2025arw,Cielo:2025ibc, Nayeri:2026ieg} among others for approaches to decoherence in cosmology).
Tracing over these unobserved environmental degrees of freedom causes the pure quantum state of the perturbations to evolve into a mixed state density matrix $\hat{\rho}$.

Current observations tightly constrain several macroscopic properties of this putative cosmological mixed state $\hat{\rho}$. 
Specifically, CMB measurements dictate that the state must be highly adiabatic, well-approximated by a Gaussian distribution, and possess a fixed variance of the amplitude of its fluctuations \cite{Planck:2018jri,Planck:2019kim}.
However, these observational constraints alone leave many kinematic properties of $\hat{\rho}$ strictly undetermined.
To fix the remaining degrees of freedom, the standard approach in the literature is to specify the pointer basis: the basis in which environmental interactions actively suppress off-diagonal quantum coherence, rendering $\hat{\rho}$ diagonal.
Several distinct proposals for the inflationary pointer basis exist, including the coherent state basis \cite{Matacz:1992tp,Campo:2004sz,Campo:2005sy}, the field configuration (amplitude) basis \cite{Kiefer:2006je, Kiefer:2008ku, Burgess:2014eoa, Nelson:2016kjm, Burgess:2022nwu}, and the particle number basis \cite{Brandenberger:1992sr,Brandenberger:1992jh,Prokopec:1992ia,Gasperini:1992xv,Gasperini:1993mq,Brahma:2020zpk}.

Rather than assuming a specific environmental interaction or restricting ourselves to a completely diagonalized mixed state, in this paper we introduce a general parameterization for any Gaussian mixed state of primordial perturbations. 
This allows us to construct a complete ``landscape" of physically allowed mixed states. 
Using this ansatz, we can systematically map the locations of the pure quantum state and the various pointer basis models from the literature, providing a unified geometric framework for evaluating decoherence.

With this landscape established,
we can evaluate the boundary of the quantum-to-classical transition.
Several theoretical diagnostics have been proposed to measure whether $\hat \rho$ retains residual quantum features or behaves purely classically, including criteria for quantum separability \cite{Campo:2005sy,Campo:2008ju,Micheli:2022tld}, quantum discord \cite{Ollivier:2001fdq,Lim:2014uea, Martin:2015qta, Martin:2021qkg,Martin:2021znx, Martin:2022kph}, the potential violation of cosmic Bell inequalities \cite{Campo:2005sv,Martin:2017zxs,Martin:2019wta,Espinosa-Portales:2022yok,Martin:2022kph}, and the quantification of non-classical resources via quantum magic \cite{Haque:2025pav,Ireland:2026txt}.
Additional observational signatures of this transition may be visible in power spectrum modifications induced by decoherence \cite{Martin:2018zbe}, and distinct analytic structures in cosmological correlators, where the existence of specific poles may serve as a unique signal of classical versus quantum evolution \cite{Green:2020whw,Ballesteros:2026evr}.
In this work, we specifically utilize the Glauber-Sudarshan $P$-representation \cite{Glauber:1963tx,Sudarshan:1963ts} as our primary diagnostic for classicality. 
By demanding that the state admits a regular, positive-definite $P$-function, we can draw a sharp threshold on our landscape, precisely delineating the mathematical regime where the state becomes indistinguishable from a classical stochastic ensemble.

Crucially, exploring this landscape reveals that decoherence carries
profound dynamical consequences for the evolution of the universe. 
In the standard inflationary narrative, the curvature perturbation $\zeta$ associated with primordial fluctuations becomes strictly constant on superhorizon scales. 
Consequently, when utilizing a classical stochastic distribution description of inflation, usually only the growing (constant) mode is retained, and the time-dependence of the curvature perturbation -- the singular decaying mode -- is set to zero \cite{Kiefer:1998jk,Campo:2004sz, Campo:2005sy}.
However, different mixed states across our landscape will generically excite this decaying mode to varying degrees due to the lack of purity of the state. 
In fact, for a state to become classical (in the sense of possessing a regular $P$-function), it must possess a momentum variance that strictly exceeds the vacuum level. 
If this momentum-driven time-dependence is sufficiently large, it threatens to spoil the precise temporal coherence of the acoustic peaks in the CMB \cite{Dodelson:2003ip} and fundamentally alter the background dynamics.

In this work, we demonstrate that the excitation of the decaying mode places stringent observational and theoretical constraints on models of cosmological decoherence. After introducing the landscape of possible mixed states consistent with observations, we determine the exact coefficient of the decaying mode produced by a generic mixed state. Finally, we analyze the severe backreaction of this time-dependent curvature perturbation on the Newtonian gravitational potential at the onset of the radiation-dominated era, establishing absolute bounds on the viability of environment-induced decoherence models.

\section{Quantum Cosmological Perturbations}
\label{sec:review_pure}

We consider scalar perturbations of a spatially flat Friedmann-Lemaître-Robertson-Walker (FLRW) background metric during inflation
\begin{equation}
    ds^2 = -dt^2 + a(t)^2 d\vec x^2\, .
\end{equation}
The expansion rate of the background is characterized by the Hubble rate $H = \dot a/a$. We are primarily interested in slow-roll accelerated expansion, where the Hubble rate is approximately constant, parameterized by the slow-roll parameters $\epsilon = -\dot H/H^2 \ll 1$ and $\eta = \dot \epsilon/(H\epsilon)$ with $|\eta|\ll 1$.
It is important to note that this strict slow-roll assumption explicitly  excludes transient non-attractor phases, such as ultra-slow-roll (USR), where $\eta \sim {\mathcal O}(1)$ \cite{Kinney:2005vj}.

In the absence of anisotropic stress, the scalar metric perturbations can be written in conformal time $\eta$ as
\begin{equation}
    ds^2 = a(\eta)^2\left(-(1+2\Phi(\eta,\vec x))d\eta^2 + (1-2\Phi(\eta,\vec{x}))d\vec x^2\right)\, .
\end{equation}
These metric perturbations combine with perturbations of the inflaton scalar field $\varphi(\eta,\vec x) = \varphi_0(\eta) + \delta \varphi(\eta,\vec x)$ to form the gauge-invariant curvature perturbation $\zeta = \Phi + (H/\dot \varphi)\delta \varphi$ \cite{Sasaki:1986hm,Mukhanov:1988jd,Mukhanov}. 
To proceed with quantization, we transform to the canonically normalized Mukhanov variable $v(\eta,\vec x) = M_{pl}a\sqrt{2\epsilon}\ \zeta $.
In terms of this variable, the quadratic action takes the form
\begin{equation}
    S_2 = \frac{1}{2} \int d\eta d^3x \left[ v'^2 - (\partial_i v)^2 + \frac{z''}{z} v^2 \right]\,,
\end{equation}
where primes denote derivatives with respect to conformal time and $z \equiv aM_{pl}\sqrt{2\epsilon}$.

Promoting the perturbation to a quantum field and expanding in Fourier modes
\begin{equation}
    \hat v(\eta,\vec x) = \int \frac{d^3k}{(2\pi)^{3/2}} \hat v_{\mathbf k}(\eta) e^{i\mathbf k\cdot \mathbf x}\,,
\end{equation}
we can define the mode operator and its canonical momentum in terms of creation and annihilation operators $\hat c_{\mathbf k}$ and $\hat c_{\mathbf k}^\dagger$
\begin{equation}
    \hat v_{\mathbf k} = \frac{1}{\sqrt{2k}}\left(\hat c_{\mathbf k} + \hat c_{-\mathbf k}^\dagger\right), \qquad \hat p_{\mathbf k} = -i\sqrt{\frac{k}{2}} \left(\hat c_{\mathbf k} - \hat c_{-\mathbf k}^\dagger\right)\, .
    \label{eq:CreationAnnihilationDefn}
\end{equation}
Because $\hat v(\eta,\vec{x})$ is real, the Fourier modes satisfy the reality condition $\hat v_{\mathbf k}^\dagger = \hat v_{-\mathbf k}$. 
The isotropy of the background ensures that the modes decouple into independent pairs with opposite momenta $(\mathbf{k}, -\mathbf{k})$. The Hamiltonian for each mode pair is then
\begin{equation}
    \hat{H} = \int d^3k \left[ k \left( \hat{c}_{\mathbf{k}}^\dagger \hat{c}_{\mathbf{k}} + \hat{c}_{-\mathbf{k}}^\dagger \hat{c}_{-\mathbf{k}} + 1 \right) - i \frac{z'}{z} \left( \hat{c}_{\mathbf{k}} \hat{c}_{-\mathbf{k}} - \hat{c}_{\mathbf{k}}^\dagger \hat{c}_{-\mathbf{k}}^\dagger \right) \right]\,.
\end{equation}
The first term dictates the free evolution of the modes, while the second term, proportional to $z'/z$, acts as a time-dependent pump that creates entangled $(\mathbf{k},-\mathbf{k})$ particle pairs due to the expanding background.
To connect the phase space dynamics directly to cosmological observables, it is useful to relate the canonical momentum $\hat{p}_{\mathbf{k}}$ to the time-evolution of the curvature perturbation $\hat{\zeta}_{\mathbf{k}}$.
Starting from the definition of the Fourier mode Mukhanov variable $\hat{v}_{\mathbf{k}} = z(\eta) \hat{\zeta}_{\mathbf{k}}$, the conjugate momentum derived from the quadratic action as 
\begin{equation}
\hat{p}_{\mathbf{k}} = \hat{v}_{\mathbf{k}}' - \frac{z'}{z} \hat{v}_{\mathbf{k}} = z(\eta) \hat{\zeta}_{\mathbf{k}}'\, .
\label{eq:MukhanovMomentum}
\end{equation} 
Therefore, the conjugate momentum directly tracks the conformal time derivative of the comoving curvature perturbation.

Assuming the modes originate in the two-mode Bunch-Davies vacuum in the far past ($|k\eta_i| \gg 1$), defined by $\hat c_{\mathbf k}(\eta_i) |0_{\mathbf k},0_{-\mathbf k}\rangle = 0$ , the unitary time-evolution factorizes into a sequence of two-mode rotation and squeezing operations \cite{Schumaker:1986tlu,PhysRevD.42.3413,PhysRevD.50.4807}
\begin{equation}
    \hat {\mathcal U}_{\mathbf k} = \hat {\mathcal S}_{\mathbf k}(r_k,\phi_k)\hat {\mathcal R}_{\mathbf k}(\theta_k)\, ,
    \label{eq:CosmoUnitary}
\end{equation}
where $\hat {\mathcal R}_{\mathbf k}(\theta_k)$ is the two-mode rotation operator
\begin{equation}
    \hat {\mathcal R}_{\mathbf k}(\theta_k) \equiv \exp\left[-i\theta_k(\eta) (\hat c_{\mathbf k}(\eta_i)\hat c_{\mathbf k}^\dagger(\eta_i) + \hat c_{-\mathbf k}^\dagger(\eta_i)\hat c_{-\mathbf k}(\eta_i))\right]\, ,
\end{equation}
and $\hat {\mathcal S}_{\mathbf k}(r_k,\phi_k)$ is the two-mode squeezing operator
\begin{equation}
    \hat {\mathcal S}_{\mathbf k}(r_k,\phi_k) \equiv \exp\left[\frac{r_k(\eta)}{2}\left(e^{-2i\phi_k(\eta)}\hat c_{\mathbf k}(\eta_i)\hat c_{-\mathbf k}(\eta_i) - e^{2i\phi_k(\eta)}\hat c_{-\mathbf k}^\dagger(\eta_i)\hat c_{\mathbf k}^\dagger(\eta_i)\right)\right]\, .
\end{equation}
This evolution transforms the vacuum into a highly entangled two-mode squeezed state 
\begin{equation}
    |\Psi\rangle_{\mathbf k,-\mathbf k} = \frac{e^{-i\theta_k}}{\cosh r_k}\sum_{n=0}^\infty (-1)^n e^{-2in\phi_k} \tanh^n \left(r_k\right) |n_{\mathbf k},n_{-\mathbf k}\rangle\,.
    \label{eq:PureSqueezed}
\end{equation}
The evolution of the squeezing parameter $r_k(\eta)$, the squeezing angle $\phi_k(\eta)$, and the rotation angle $\theta_k(\eta)$ are governed by coupled equations of motion dependent on the background pump parameter $z'/z$ \cite{PhysRevD.42.3413,PhysRevD.50.4807}.
The rotation angle $\theta_k$ will not play a role in our analysis, so we will drop it.

The two-mode squeezed state \eqref{eq:PureSqueezed} is Gaussian, and is characterized by its two-point functions
\begin{align}
\label{eq:PureOccupation}
\langle \hat c_{\mathbf{k}}^\dagger \hat c_{\mathbf{k}'}\rangle &= \sinh^2 r_k\, (2\pi)^3 (2\pi)^3 \delta^3(\mathbf{k} - \mathbf{k}')\, ;\\[4pt]
\label{eq:PureCoherence}
\langle \hat c_{\mathbf{k}} \hat c_{\mathbf{k}'}\rangle &= -\tfrac{1}{2}\,e^{-2 i\phi_k}\sinh(2r_k)\, (2\pi)^3 \delta^3(\mathbf{k} + \mathbf{k}')\, ;\qquad(\text{with}\;\langle \hat c_{\mathbf{k}}^\dagger \hat c_{\mathbf{k}'}^\dagger\rangle=\langle \hat c_{\mathbf{k}} \hat c_{\mathbf{k}'}\rangle^\dagger).
\end{align}
with all other two-point functions vanishing.
When translated into the Mukhanov variables, these become
\begin{subequations} \label{eq:Pure2Pt}
\begin{align}
    \langle \hat v_{\mathbf{k}}\hat v_{\mathbf{k}'}\rangle
&\equiv\sigma_v^2\, (2\pi)^3\delta^3(\mathbf{k}+\mathbf{k}') = \frac{1}{2k}\!\left[\cosh2r_k-\cos(2\phi_k)\sinh2r_k\right](2\pi)^3\delta^3(\mathbf{k}+\mathbf{k}')\, ; \label{eq:Pure2Pt:1} \\
\langle \hat p_{\mathbf{k}}\hat p_{\mathbf{k}'}\rangle
&\equiv\sigma_p^2\, (2\pi)^3\delta^3(\mathbf{k}+\mathbf{k}')=\frac{k}{2}\!\left[\cosh2r_k+\cos(2\phi_k)\sinh2r_k\right] (2\pi)^3\delta^3(\mathbf{k}+\mathbf{k}')\, ; \label{eq:Pure2Pt:2}\\
\frac12\langle\{\hat v_{\mathbf{k}},\hat p_{\mathbf{k}'}\}\rangle
&\equiv\sigma_{vp}\, (2\pi)^3\delta^3(\mathbf{k}+\mathbf{k}')=\frac{1}{2}\sinh2r_k\sin(2\phi_k)\, (2\pi)^3\delta^3(\mathbf{k}+\mathbf{k}')\, .\label{eq:Pure2Pt:3}
\end{align}
\end{subequations}
To fully characterize the quantum state of these perturbations in phase space, we should construct the corresponding density matrix and Wigner function.
From \eqref{eq:PureSqueezed}, the density matrix in number basis is 
\begin{align}
    \hat\rho_{\mathbf{k},\rm pure} &= |\Psi\rangle_{\mathbf{k},-\mathbf{k}}\langle\Psi|_{\mathbf{k},-\mathbf{k}} \\
    &= \frac{1}{\cosh^2 r_k} \sum_{n=0}^{\infty} \sum_{m=0}^{\infty}  (-1)^{m+n} e^{-2i\phi_k(m-n)}(\tanh r_k)^{n+m}
    |n_{\mathbf{k}}, n_{-\mathbf{k}}\rangle \langle m_{\mathbf{k}}, m_{-\mathbf{k}}|\,.
    \label{eq:2ModeSqueezedStateDensity}
\end{align}
Notice that in number basis, the density matrix is not diagonal, indicating the presence of quantum correlations between the $(\mathbf{k},-\mathbf{k})$ modes.
Because the original Fourier mode operators $\hat{v}_{\mathbf{k}}$ and $\hat{p}_{\mathbf{k}}$ are non-Hermitian, satisfying $\hat v_{\mathbf{k}}^\dagger = \hat v_{-\mathbf{k}}$ (thus  mixing the $\mathbf{k}$ and $-\mathbf{k}$ sectors), we will transform to a set of real, Hermitian variables in order to construct a ``position-space" representation of the state.

There are two common choices of Hermitian basis for cosmological perturbations.
The first approach we will consider uses the Hermitian position-like and momentum-like variables for the individual $\mathbf{k}$ and $-\mathbf{k}$ harmonic oscillators:
\begin{align}
\hat X_{\mathbf{k}} = \frac{1}{\sqrt{2k}} \left(\hat c_{\mathbf{k}} + \hat c_{\mathbf{k}}^\dagger\right)\,, \qquad
\hat P_{\mathbf{k}} = -i\sqrt{\frac{k}{2}} \left(\hat c_{\mathbf{k}} - \hat c_{\mathbf{k}}^\dagger\right)\,,
\end{align}
These variables are advantageous because they do not mix the creation and annihilation operators of the $\mathbf{k}$ and $-\mathbf{k}$ modes. 
However, the price paid is that the Hamiltonian couples the opposite momenta through terms such as $\hat X_{\mathbf{k}} \hat P_{-\mathbf{k}}$.

Alternatively, we can decompose the Mukhanov variable and its canonical momentum directly into their ``real" and ``imaginary" components \cite{Martin:2015qta}
\begin{align}
\hat v_{\mathbf{k}} = \frac{1}{\sqrt{2}} \left(\hat q_{\mathbf{k}}^R + i \hat q_{\mathbf{k}}^I\right), \qquad
\hat p_{\mathbf{k}} = \frac{1}{\sqrt{2}} \left(\hat p_{\mathbf{k}}^R + i \hat p_{\mathbf{k}}^I\right)\,,
\label{eq:RealImaginaryVariables}
\end{align}
where the new operators $\hat q_{\mathbf{k}}^{R,I}$ and $\hat p_{\mathbf{k}}^{R,I}$ are manifestly Hermitian. The reality condition $\hat v_{\mathbf{k}}^\dagger = \hat v_{-\mathbf{k}}$ imposes parity constraints such that $(\hat q_{\mathbf{k}}^R)^\dagger = \hat q_{-\mathbf{k}}^R = \hat q_{\mathbf{k}}^R$ (parity-even) and $(\hat q_{\mathbf{k}}^I)^\dagger = -\hat q_{-\mathbf{k}}^I = \hat q_{\mathbf{k}}^I$ (parity-odd). 
To avoid double-counting degrees of freedom, the momentum for this basis will be restricted to half-space $\mathbb{R}^{3+}$. 
A primary advantage of this decomposition is that the Hamiltonian decouples the real and imaginary parts from each other entirely.
We will primarily work in this basis for our analysis, though naturally it is possible to work in the other basis.

Working in the $(q_{\mathbf{k}}^R, q_{\mathbf{k}}^I)$ amplitude basis, we can construct the density matrix for the pure two-mode squeezed state. Using the property of the pure state $\hat{\rho}_{\mathbf{k}} = |\Psi\rangle_{\mathbf{k},-\mathbf{k}}\langle\Psi|_{\mathbf{k},-\mathbf{k}}$, the density matrix elements in this basis take the form of a Gaussian:
\begin{align}
\rho_{\rm pure}(q_{\mathbf{k}}^R, q_{\mathbf{k}}^I; \tilde q_{\mathbf{k}}^{R}, \tilde q_{\mathbf{k}}^{I}) = 
\frac{1}{2\pi\sigma_v^2} \exp{\Bigg(}&-\frac{\left[(q_{\mathbf{k}}^R+\tilde q_{\mathbf{k}}^R)^2 + (q_{\mathbf{k}}^I+\tilde q_{\mathbf{k}}^I)^2\right]}{8\sigma_v^2} - \frac{\left[(q_{\mathbf{k}}^R-\tilde q_{\mathbf{k}}^R)^2 + (q_{\mathbf{k}}^I-\tilde q_{\mathbf{k}}^I)^2\right]}{8\sigma_v^2}  \nonumber \\
&- i\frac{\sigma_{vp}}{2\sigma_v^2}\left[((q_{\mathbf{k}}^R)^2-(\tilde q_{\mathbf{k}}^R)^2) + ((q_{\mathbf{k}}^I)^2-(\tilde q_{\mathbf{k}}^I)^2)\right] {\Bigg )}\,.
\label{eq:RIDensityMatrix}
\end{align}
For a pure state, the spread of the off-diagonal terms is correlated with the diagonal width through $\sigma_v^2$.

A useful geometric way to visualize the state is expressed using the Wigner function $W(\mathbf{r})$ \cite{Polarski:1995jg}, which serves as a quasi-probability distribution in the phase space $\mathbf{r}^T = (q_{\mathbf{k}}^R, p_{\mathbf{k}}^R, q_{\mathbf{k}}^I, p_{\mathbf{k}}^I)$. 
In general, a two-mode Gaussian Wigner function using these variables 
decouples into a product of identical real and imaginary sector distributions:
\begin{equation}
    W(\mathbf{r}) = \frac{1}{(2\pi)^2 \sqrt{\det \mathbf{\Sigma}}} \exp\left[-\frac{1}{2} \mathbf{r}^T \mathbf{\Sigma}^{-1} \mathbf{r}\right]\,.
    \label{eq:GeneralTwoModeWigner}
\end{equation}
The full $4 \times 4$ covariance matrix $\mathbf{\Sigma}$ takes a block-diagonal form:
\begin{equation}
    \mathbf{\Sigma} = \begin{pmatrix}\mathbf{\Sigma}_1 &  0 \cr
0 & \mathbf{\Sigma}_1
\end{pmatrix}, 
\hspace{.2in} \text{where} \hspace{.2in} \mathbf{\Sigma}_1 = \begin{pmatrix}
\sigma_v^2 & \sigma_{vp} \cr
\sigma_{vp} & \sigma_p^2
    \end{pmatrix}\,.
\label{eq:RISigma}
\end{equation}
Because of assumed spatial isotropy and momentum conservation, cross-correlators between the real and imaginary sectors for the pure state (such as $\langle \hat q_{\mathbf{k}}^R \hat p_{\mathbf{k}}^I\rangle_{\rm pure}=0$) vanish, as they pair parity-even and parity-odd operators. 
Notice that because the state generated by standard inflationary dynamics is a pure squeezed state, the determinant of the sub-matrix is minimal, $\det \mathbf{\Sigma}_1 = \sigma_v^2 \sigma_p^2 - \sigma_{vp}^2 = 1/4$, saturating the Heisenberg uncertainty relation. 

\begin{figure}[t]
\centering\includegraphics[width=0.6\textwidth]{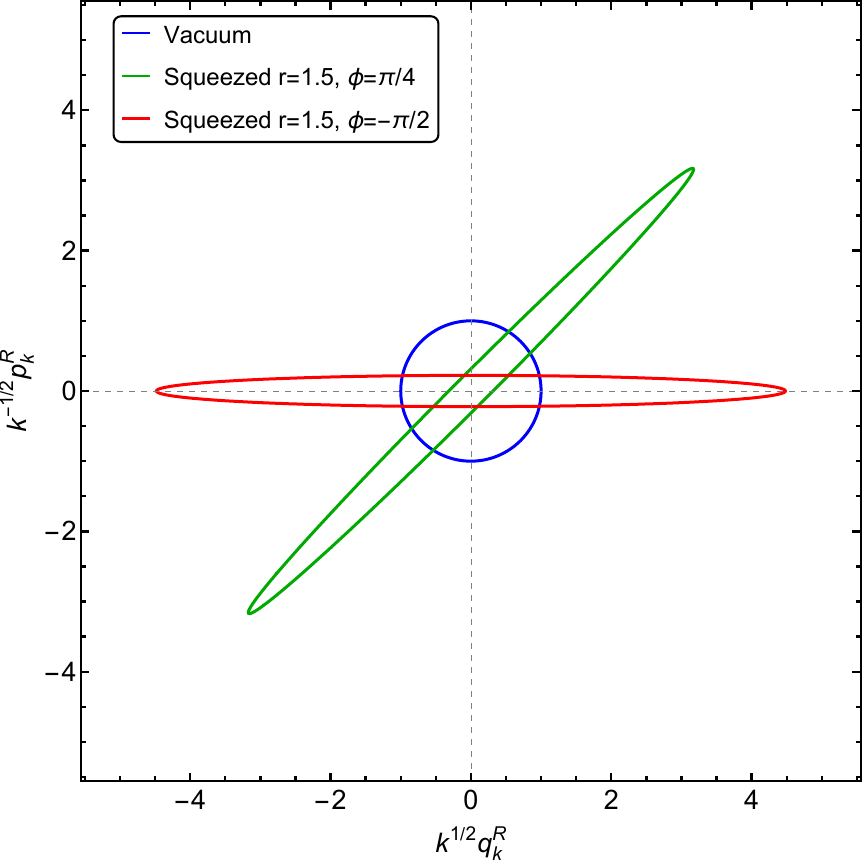}
\caption{The 1-$\sigma$ contours of the Wigner function $W(k^{1/2} q_{\mathbf{k}}^R,k^{-1/2}p_{\mathbf{k}}^R,0,0)$ for a vacuum state (blue contour), a general squeezed state (green), and the long-wavelength squeezing induced by inflation which dynamically sets the squeezing angle $\phi = -\pi/2$ (red) in a squeezed-momentum configuration.}
\label{fig:WignerPure}
\end{figure}

Geometrically, the Wigner function provides a powerful visualization of the state in phase space.
For the vacuum state $r_k = 0$, the Wigner function is an isotropic Gaussian, as in Figure \ref{fig:WignerPure}.
This circular distribution represents the isotropic vacuum fluctuations of the initial ground state.
For a squeezed state with squeezing parameters $r_k, \phi_k$, the Wigner function is an ellipse whose semi-major axes are controlled by the squeezing parameter $r_k$ and orientation is controlled by the squeezing angle $\phi_k$, as in Figure \ref{fig:WignerPure}.

During inflation, the background expansion is approximately de Sitter, with the scale factor evolving as $a(\eta) = -1/(H_{dS}\, \eta)$ for $-\infty < \eta < 0$. In this inflationary limit, the squeezing and rotation parameters admit the solutions \cite{PhysRevD.42.3413,PhysRevD.50.4807}
\begin{subequations}\label{eq:dSSolution}
\begin{align}
r_k &= \sinh^{-1} \left(\frac{1}{2k|\eta|}\right)\, ; \label{eq:dSSolution:1} \\
\phi_k &= -\frac{\pi}{4} - \frac{1}{2} \tan^{-1} \left(\frac{1}{2k|\eta|}\right)\, ; \label{eq:dSSolution:2} \\
\theta_k & = k|\eta| - \tan^{-1}\left(\frac{1}{2k|\eta|}\right) \, .\label{eq:dSSolution:3}
\end{align}
\end{subequations}
At early times when the modes are deep inside the horizon ($k|\eta| \gg 1$), the squeezing is negligible ($r_k \ll 1$), recovering the standard vacuum state. 
However, at late times as the modes exit the horizon ($k|\eta| \ll 1$), the squeezing parameter grows proportionally to the number of e-folds since horizon exit, $r_k \sim N_e = \ln a/a_e \gg 1$, while the squeezing angle asymptotes to the constants $\phi_k \approx -\pi/2$ (the rotation angle $\theta_k$ will not be relevant for our analysis, so we will drop it from our consideration in the following).

For the solutions \eqref{eq:dSSolution:1}-\eqref{eq:dSSolution:3},
the two-point functions \eqref{eq:Pure2Pt:1}-\eqref{eq:Pure2Pt:3} for the Mukhanov variables become the exact results
\begin{subequations}
    \begin{align} 
\sigma_{v,\text{inf}}^2 &= \frac{1}{2k} \left(1+\frac{1}{(k\eta)^2}\right)= \frac{1}{2k} \left(e^{2r_k}-1+e^{-2r_k}\right) \,; \label{eq:Pure2PtdS:1}\\
\sigma_{p,\text{inf}}^2 &= \frac{k}{2} = \sigma_{p,\rm vac}^2\,;\label{eq:Pure2PtdS:2}\\
\sigma_{vp,\text{inf}} &= \frac{1}{2} \frac{1}{k|\eta|}= \frac{1}{2} \left(e^{r_k} - e^{-r_k}\right) \, .\label{eq:Pure2PtdS:3}
\end{align}
\end{subequations}
In the superhorizon limit, the Wigner function phase space contour is thus vastly stretched along the amplitude axis and tightly pinched along the momentum axis, as in Figure \ref{fig:WignerPure}. For $N_e \sim 60$ e-folds of inflation, the extreme squeezing leads to strongly enhanced amplitude fluctuations $\sigma_{v,\rm inf}^2 \sim e^{60}$.
This highly squeezed ellipse visually captures the physical mechanism of inflation: the generation of macroscopic, quasi-classical density perturbations through squeezing.
It is a remarkable feature of cosmological perturbation theory on a de Sitter background that the extreme exponential squeezing that stretches the amplitude quadrature does not force an equally extreme exponentially small momentum quadrature.
Instead, the time-dependence of the squeezing parameter $r_k$ and squeezing angle $\phi_k$ guarantee that the momentum variance stays fixed at its vacuum value, $\sigma_p^2 = k/2$. 
In Appendix \ref{app:MomentumVariance}, we demonstrate how the momentum operator $\hat p_{\mathbf k}$ decouples from the cosmic expansion in a de Sitter background, and show how in a realistic inflationary scenario in which the slow roll dynamics generate a quasi-de Sitter background, the momentum variance drifts slowly from the vacuum limit as $\sigma_p^2/\sigma_{p,\rm vac}^2 \approx e^{2\delta \Delta N}$, where $\delta \sim {\mathcal O}(\epsilon)$ is of the order of the slow roll parameter $\epsilon$, and $\Delta N$ is the number of e-folds for the mode since horizon crossing. For $\epsilon \sim 0.01$ and $\Delta N \sim 60$, this drift corresponds to a factor of $\sigma_p^2/\sigma_{p,\rm vac}^2 \approx 3$.

While the two-point function of the Mukhanov variable $\hat v_{\mathbf k}$ grows exponentially with squeezing, the amplitude of the co-moving curvature perturbation $\zeta_{\mathbf k}$ becomes constant and scale-invariant
\begin{equation}
    \langle \hat\zeta_{\mathbf k'} \hat\zeta_{\mathbf k}\rangle \approx \frac{e^{2r_k}}{4 k a^2(\eta) M_{pl}^2\epsilon} (2\pi)^3 \delta^3(\mathbf k+\mathbf k') = \frac{8\pi^2}{k^3} \Delta_\zeta^2\ (2\pi)^3 \delta^3(\mathbf k + \mathbf k')\,,
    \label{eq:InflationaryPowerSpectrum}
\end{equation}
where 
\begin{equation}
\Delta^2_\zeta \equiv \frac{H_{\rm dS}^2}{8\pi^2 \epsilon M_{pl}^2}\sim 2 \times 10^{-9}    
\end{equation}
is the dimensionless power spectrum set by observations \cite{Planck:2018jri}.
The two-point function for the time-derivative of the curvature perturbation $\zeta'_{\mathbf{k}}$ for an inflationary background is correspondingly suppressed on large scales
\begin{equation}
    \langle \zeta'_{\mathbf{k}} \zeta'_{\mathbf{k}'}\rangle \approx \frac{k}{4 a^2(\eta) M_{pl}^2\epsilon} \ (2\pi)^3\ \delta^3(\mathbf{k}+\mathbf{k}') = \frac{(H_{\rm dS} \eta)^2}{M_{pl}^2 \epsilon} \frac{k}{4} \ (2\pi)^3\ \delta^3(\mathbf{k}+\mathbf{k}')\,,
\end{equation}
where $\tau \rightarrow 0^-$, so that the pure state cosmological curvature perturbations are constant on superhorizon scales.

This purely Gaussian state is the baseline prediction of standard inflationary cosmology. 
Any classicalization or decoherence of the state will present as a modification to this structure, decreasing the purity of the density matrix and altering the elements of the covariance matrix.

\section{A Decoherence Landscape of Cosmological Perturbations}
\label{sec:decoherence}

\subsection{Parameterizing Gaussian Mixed States}

To properly study decoherence during inflation, we must characterize deviations from the pure state evolution described in the previous section. Because observations of the cosmic microwave background (CMB) indicate that primordial perturbations are highly Gaussian, we will restrict our analysis entirely to two-mode Gaussian mixed states.

For a real scalar field, momentum conservation requires that correlators satisfy $\langle \ldots_{\mathbf{k}} \ldots_{\mathbf{p}}\rangle \sim \delta^3(\mathbf{k}+\mathbf{p})$. This implies that only correlators pairing the $\mathbf{k}$ and $-\mathbf{k}$ modes survive. 
In addition, spatial isotropy ($\mathbf{k} \leftrightarrow -\mathbf{k}$) ensures that cross-correlators between the parity-even real parts and parity-odd imaginary parts vanish (e.g., $\langle \hat q_{\mathbf{k}}^R \hat p_{\mathbf{k}}^I\rangle = 0$). 
Consequently, the most general Gaussian Wigner function for two coupled Fourier modes is controlled by only three parameters. These parameters correspond to the amplitude variance $\sigma_v^2 = \langle |\hat v_{\mathbf{k}}|^2 \rangle$, the momentum variance $\sigma_p^2 = \langle |\hat p_{\mathbf{k}}|^2 \rangle$, and their covariance $\sigma_{vp} = \frac{1}{2} \langle \hat v_{\mathbf{k}} \hat p_{\mathbf{k}}^\dagger + \hat p_{\mathbf{k}} \hat v_{\mathbf{k}}^\dagger \rangle$.

In the phase-space basis $\mathbf{r}^T = (q_{\mathbf{k}}^R, p_{\mathbf{k}}^R, q_{\mathbf{k}}^I, p_{\mathbf{k}}^I)$, the Wigner function for this general two-mode Gaussian state takes the form
\begin{equation}
W(\mathbf{r}) = \frac{1}{(2\pi)^2 \sqrt{\det \mathbf{\Sigma}}} \exp\left[-\frac{1}{2} \mathbf{r}^T \mathbf{\Sigma}^{-1} \mathbf{r}\right]\,.
\label{eq:WignerMixed}
\end{equation}
As with the pure state in the previous section, because the real and imaginary sectors decouple due to parity
the full $4 \times 4$ covariance matrix $\mathbf{\Sigma}$ is block-diagonal, composed of two identical $2 \times 2$ sub-matrices $\mathbf{\Sigma}_1$
\begin{equation}
\mathbf{\Sigma} = \begin{pmatrix} \mathbf{\Sigma}_1 & 0 \\ 0 & \mathbf{\Sigma}_1 \end{pmatrix}, \hspace{.2in} \text{where} \hspace{.2in} \mathbf{\Sigma}_1 = \begin{pmatrix} \sigma_v^2 & \sigma_{vp} \\ \sigma_{vp} & \sigma_p^2 \end{pmatrix}\,.
\label{eq:WignerSubMatrix}
\end{equation}
The corresponding density matrix in the coordinate (amplitude) representation is
\begin{align}
\rho(q_{\mathbf{k}}^R, q_{\mathbf{k}}^I; \tilde q_{\mathbf{k}}^{R}, \tilde q_{\mathbf{k}}^{I}) = \mathcal{N} \exp\Bigg\{&-A\left[(q_{\mathbf{k}}^R+\tilde q_{\mathbf{k}}^R)^2 + (q_{\mathbf{k}}^I+\tilde q_{\mathbf{k}}^I)^2\right] - B\left[(q_{\mathbf{k}}^R-\tilde q_{\mathbf{k}}^R)^2 + (q_{\mathbf{k}}^I-\tilde q_{\mathbf{k}}^I)^2\right]  \nonumber \\
&- iD\left[((q_{\mathbf{k}}^R)^2-(\tilde q_{\mathbf{k}}^R)^2) + ((q_{\mathbf{k}}^I)^2-(\tilde q_{\mathbf{k}}^I)^2)\right] \Bigg\}\,,
\label{eq:MixedDensityMatrixq}
\end{align}
where the coefficients are given directly by the elements of the covariance matrix: $\mathcal{N} = \frac{1}{2\pi\sigma_v^2}$, $A = \frac{1}{8\sigma_v^2}$, $B = \frac{\sigma_v^2 \sigma_p^2 - \sigma_{vp}^2}{2\sigma_v^2}$, and $D = \frac{\sigma_{vp}}{2\sigma_v^2}$.
Notice that the primary difference with the pure state expression \eqref{eq:RIDensityMatrix} is that the coefficient $B$ in \eqref{eq:MixedDensityMatrixq} is no longer identical to the coefficient $A$.

To quantify the degree to which the state has decohered from a pure quantum state into a statistical mixture, we analyze the purity of the state, defined generally as $\mu = \text{Tr}(\hat \rho^2)$. For a pure state, the density matrix is idempotent ($\hat \rho^2 = \hat \rho$), yielding $\mu = 1$
For Gaussian states, the purity can be expressed entirely in terms of the determinant of the covariance matrix. Focusing on the $2 \times 2$ sub-matrix $\mathbf{\Sigma}_1$, the purity parameter takes the form
\begin{equation}
    \mu = \frac{1}{4 \det \mathbf{\Sigma}_1} = \frac{1}{4(\sigma_v^2 \sigma_p^2 - \sigma_{vp}^2)}\,.
    \label{eq:PurityCovarianceDefn}
\end{equation}
The Heisenberg uncertainty principle mandates that $\det \mathbf{\Sigma}_1 = \sigma_v^2 \sigma_p^2 - \sigma_{vp}^2 \ge 1/4$, which restricts the purity to the physical range $0 < \mu \leq 1$.
If the state is a pure state, such as the two-mode squeezed state with variances given by \eqref{eq:Pure2Pt:1}-\eqref{eq:Pure2Pt:3}, the uncertainty bound is saturated ($\det \mathbf{\Sigma}_1 = 1/4$) and the purity is maximal $\mu=1$.
However, for mixed states we have $\det \mathbf{\Sigma}_1 > 1/4$ so that the purity is strictly less than one ($\mu < 1$).
While we choose to parameterize the mixed state using the purity, for Gaussian 2-mode states this is equivalent to specifying the von Neumann  entropy $S = -\text{Tr}(\hat \rho_{\rm r} \ln \hat \rho_{\rm r})$, which
has historically been a standard measure of cosmological decoherence \cite{Brandenberger:1992sr,Brandenberger:1992jh,Prokopec:1992ia,Gasperini:1992xv,Gasperini:1993mq,Kiefer:1998jk,Kiefer:1998pb,Kiefer:1998qe,Campo:2004sz,Brahma:2020zpk}.

We can now map out the landscape of physically allowed Gaussian pure and mixed states. 
Crucially, any valid model of decoherence must remain consistent with cosmic microwave background (CMB) observations. 
The CMB precisely fixes the amplitude of the comoving curvature perturbations, effectively anchoring the amplitude two-point function $\langle |\hat{v}_{\mathbf{k}}|^2 \rangle = \sigma_v^2$. 
For superhorizon modes relevant to CMB scales this variance is well-approximated by \eqref{eq:Pure2PtdS:1} in the limit of large squeezing \cite{Polarski:1995jg}
\begin{equation}
\sigma_v^2 \approx \frac{e^{2r_k}}{2k}\,.
\label{eq:FixedAmplitudeVariance}
\end{equation}
While $r_k$ is formally a time- and wavelength-dependent function $r_k(\eta)$, we evaluate it for large scale modes that have undergone  $N_e \sim 60$ $e$-folds of inflation, so a typical superhorizon squeezing parameter is $r_k \sim 60$.
By holding $\sigma_v^2$ fixed to match observations, the general two-mode Gaussian mixed state is uniquely parameterized by the two remaining degrees of freedom in the covariance matrix: the momentum variance $\sigma_p^2$ and the covariance $\sigma_{vp}$. 
However, to better interpret the physical effects of decoherence, we will trade $\sigma_{vp}$ for the state purity $\mu$. 
Furthermore, we will work with the dimensionless combination $\sigma_p^2/\sigma_{p,{\rm vac}}^2$ where $\sigma_{p,{\rm vac}}^2 = k/2$. 
This allows us to define a two-dimensional parameter space -- a ``landscape" of allowed states -- spanned by $(\sigma_p^2/\sigma_{p,{\rm vac}}^2, \mu)$.

\begin{figure}[t]
\centering\includegraphics[width=.9\textwidth]{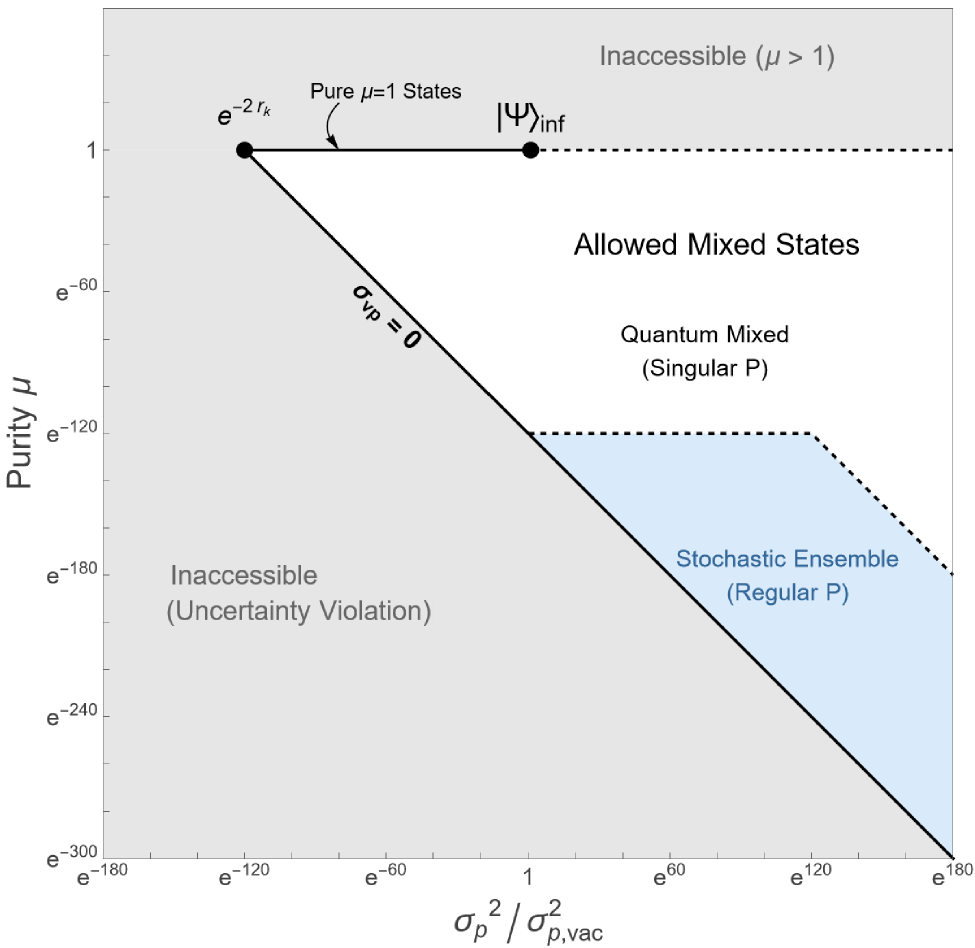}
\caption{The landscape of states describing cosmological perturbations is controlled by two dimensionless parameters (after fixing the amplitude by observations \eqref{eq:FixedAmplitudeVariance}), which we are considering here to be the purity $\mu$ and the variance of the Mukhanov momentum relative to the vacuum value $\sigma_p^2/\sigma_{p,{\rm vac}}^2$ where $\sigma_{p,{\rm vac}}^2 = k/2$. The parameter space is bounded from above by pure states so that $\mu \leq 1$ and is bounded from below by the uncertainty principle, which becomes $\mu \geq 1/(4\sigma_v^2\sigma_p^2) \sim e^{-2r_k}/(\sigma_p^2/\sigma_{p,{\rm vac}}^2)$ (a line of slope -1 on this log-log plot). The pure two-mode squeezed state in an inflationary background $|\Psi\rangle_{\text{inf}}$ is located on the $\mu=1$ boundary with a momentum variance equal to that of the vacuum.
}
\label{fig:LandscapePlotEmpty}
\end{figure}

Let us return briefly to the pure 2-mode squeezed state solutions on the inflationary background \eqref{eq:dSSolution:1}-\eqref{eq:dSSolution:3}. 
Because the state is pure, it lies on the $\mu = 1$ line; we also see that the momentum variance is fixed at the vacuum value \eqref{eq:dSSolution:2}, so the pure two-mode inflationary state is located at $(\sigma_p^2/\sigma_{p,\rm vac}^2, \mu) = (1,1)$ on Figure \ref{fig:LandscapePlotEmpty}.

Historically, 
the proposal of ``decoherence without decoherence'' \cite{Polarski:1995jg,Kiefer:2008ku} 
has been used to argue that this pure state can be treated as a classical stochastic ensemble.
This argument relies on keeping only
the leading-order ``growing mode" behavior of \eqref{eq:Pure2PtdS:1}-\eqref{eq:Pure2PtdS:3}, neglecting the ``decaying mode."
However, we can quickly see that this truncation destroys the quantum state entirely. 
By decomposing the complex mode functions into real and imaginary parts (representing the decaying mode $\alpha_k$ and the growing mode $\beta_k$), the exact covariance determinant is strictly enforced by the canonical Wronskian: $\det(\mathbf{\Sigma}) = (\alpha_k' \beta_k - \alpha_k \beta_k')^2 = 1/4$.
The ``decoherence without decoherence" approximation explicitly discards the decaying sector ($\alpha_k \to 0, \alpha_k' \to 0$).
Under this truncation, the variances collapse to $\sigma_v^2 \to \beta_k^2 \approx e^{2r_k}/(2k)$, $\sigma_p^2 \to (\beta_k')^2 \approx k/2$, and $\sigma_{vp} \to \beta_k\beta_k' \approx e^{r_k}/2$, yielding 
\begin{equation} 
\sigma_v^2 \sigma_p^2 - \sigma_{vp}^2 = 0\, . 
\end{equation}
This mathematically violates the Heisenberg uncertainty principle and the fundamental purity bound (since $\det(\mathbf{\Sigma}) = 0$ formally implies $\mu \rightarrow \infty$). 
Thus, despite its name,
``decoherence without decoherence" does not actually correspond to decoherence into a Gaussian mixed state.

More generally, since the purity cannot exceed unity the landscape is bounded from above by the line $\mu = 1$. 
Moving along this boundary corresponds to pure states with varying momentum variances.
However, even for a pure state ($\mu = 1$), the dimensionless momentum variance cannot be arbitrarily small or arbitrarily large. Because $\sigma_{vp}^2 \ge 0$, the determinant condition $\sigma_v^2 \sigma_p^2 - \sigma_{vp}^2 = 1/4$ implies $4\sigma_v^2 \sigma_p^2 \ge 1$. Writing this in terms of our dimensionless parameter yields a strict lower bound on the momentum variance
\begin{equation}
\frac{\sigma_p^2}{\sigma_{p,{\rm vac}}^2} \ge \frac{1}{2k\sigma_v^2} \approx e^{-2r_k}\, .
\end{equation}
Here we see a critical physical constraint: because the amplitude variance $\sigma_v^2$ is finite, it is mathematically impossible to access a state with exactly $\sigma_p^2 = 0$.
As we will discuss in more detail in Section 4, this strict lower bound is in direct tension with phenomenological models of cosmological perturbations (e.g., standard initialization of Boltzmann codes), which routinely set the gravitational decaying mode -- and therefore the momentum variance --identically to zero.
In our framework, we will locate this phenomenological assumption not at zero, but as saturating this vanishingly small lower bound in the inaccessible region of parameter space, $\sigma_p^2/\sigma_{p,\rm vac}^2 \rightarrow e^{-2r_k} \sim e^{-120}$.

The momentum variance for a pure state is also bounded from above.
For fixed squeezing $r_k$ the two-mode squeezed state variances \eqref{eq:Pure2Pt:1}-\eqref{eq:Pure2Pt:3} are strictly bounded, e.g. $\sigma_{vp} \leq \sinh(2r_k)/2$. For fixed variance of the amplitude \eqref{eq:FixedAmplitudeVariance}, this translates to an upper bound on the momentum variance 
\begin{equation}
    \frac{\sigma_p^2}{\sigma_{p,\rm vac}^2} < 1 + e^{-2r_k}\, .
\end{equation}
Interestingly, the momentum variance of the pure two-mode squeezed state in an inflationary background \eqref{eq:Pure2PtdS:2} comes close to saturating this bound.
The allowed parameter space for pure states is shown in Figure \ref{fig:LandscapePlotEmpty} as a solid line at $\mu = 1$.

Moving away from the pure state boundary, for any given $\sigma_p^2$, the purity cannot be made arbitrarily small. 
The state reaches its minimum possible purity when the covariance vanishes entirely ($\sigma_{vp} = 0$). In this limit, the purity is bounded by $\mu \geq 1/(4\sigma_v^2 \sigma_p^2)$. 
Substituting the fixed CMB amplitude $\sigma_v^2$, we find the lower boundary of the landscape
\begin{equation}
\mu \geq \frac{e^{-2r_k}}{(\sigma_p^2/\sigma_{p,{\rm vac}}^2)}\, .
\end{equation}
The interior region enclosed by these strict mathematical boundaries defines the complete parameter space of allowed two-mode Gaussian mixed states.
Because the amplitude variance is exponentially sensitive to the squeezing parameter ($e^{-2r_k} \sim e^{-120}$), this parameter space spans many orders of magnitude. Therefore, it is most natural to visualize this landscape on a logarithmic scale. 
On a log-log plot of $\mu$ versus $\sigma_p^2/\sigma_{p,{\rm vac}}^2$, the pure state boundary $\mu=1$ is a horizontal ceiling, while the lower boundary ($\sigma_{vp} = 0$) appears as a straight line with a slope of $-1$.
Figure \ref{fig:LandscapePlotEmpty} illustrates the main features of the allowed landscape of two-mode Gaussian mixed states describing decohered cosmological perturbations.

\subsection{Classicality and the Landscape}
\label{subsec:ClassicalPFunction}

Beyond the overall purity $\mu$, defining the precise boundary of the quantum-to-classical transition requires a robust metric for measuring the ``quantumness'' of the state. 
For example, consider \emph{Quantum Discord} ${\mathcal D}$ \cite{Ollivier:2001fdq}, which measures the broader presence of fundamental quantum correlations that cannot be accessed via classical local measurements.
While this quantity does track the degree to which a system departs from a purely classical ensemble, relying on it as an absolute criterion for classicality introduces a significant ambiguity: it is strictly dependent on how the system is partitioned into subsystems. 
For example, when the pure two-mode squeezed state \eqref{eq:2ModeSqueezedStateDensity} is partitioned into the dynamically entangled $(\mathbf k,-\mathbf k)$ momentum sectors, quantum discord is non-zero, scaling as ${\mathcal D} \sim 2r_k/\ln 2$ \cite{Lim:2014uea, Martin:2015qta, Martin:2021znx, Martin:2022kph}. 
Conversely, if the exact same state is partitioned into the real and imaginary standing-wave variables $\hat q_{\mathbf k}^R,\hat q_{\mathbf k}^I$ \eqref{eq:RealImaginaryVariables}, the density matrix factorizes and the discord vanishes identically \cite{Martin:2015qta}.
A similar criterion is the violation of cosmological Bell state inequalities  \cite{Campo:2005sv,Martin:2017zxs,Martin:2019wta,
Espinosa-Portales:2022yok,Martin:2022kph}, indicating the presence of quantum entanglement, which also have a dependence on the subsystem partitioning. 

To avoid the ambiguity of subsystem partitioning, one must look to global phase-space representations to determine whether the physical state itself is genuinely equivalent to a classical stochastic distribution. 
Historically, it has been widely argued that the highly squeezed nature of the inflationary vacuum renders the state practically indistinguishable from a classical ensemble \cite{Polarski:1995jg, Lesgourgues:1996jc, Kiefer:1998qe, Kiefer:1998jk, Kiefer:1998pb, Kiefer:2006je, Kiefer:2008ku}. 
Because the Wigner function \eqref{eq:WignerMixed} for any Gaussian state, including pure squeezed states, is everywhere positive, it is mathematically possible to treat it as a classical probability density and sample classical phase-space trajectories.
However, 
this pragmatic approach alone may not be enough to establish true classicality. The Wigner covariance matrix $\mathbf{\Sigma}_1$ inherently incorporates the irreducible zero-point fluctuations of the quantum vacuum. 
Treating the positive Wigner function as a classical stochastic distribution effectively masks mandatory Heisenberg quantum uncertainty, treating it as if it were simply classical statistical noise. 
Indeed, relying solely on the positivity of the Wigner function would classify the pure, maximally quantum initial state of inflation as classical, despite it retaining quantum discord \cite{Martin:2015qta}.

One proposal for a rigorous, unambiguous criterion for classicality requires that a distribution remain mathematically well-behaved \emph{after} the fundamental quantum vacuum fluctuations have been explicitly removed.
This threshold is naturally defined by the Glauber-Sudarshan P-representation \cite{Glauber:1963tx,Sudarshan:1963ts}, which expresses any bipartite quantum state $\hat \rho$ as a diagonal mixture of two-mode coherent states 
\begin{equation}
    \hat{\rho} = \int d^2\alpha d^2 \beta\ P(\alpha,\beta) \left(|\alpha_{\mathbf k}\rangle\langle\alpha_{\mathbf k}|\right)\otimes \left(|\beta_{-\mathbf k}\rangle\langle\beta_{-\mathbf k}|\right)\,.
\end{equation}
The $P(\alpha,\beta)$-function can be rigorously interpreted as a classical probability distribution if and only if it is a regular, non-negative function \cite{Glauber:1963tx,Sudarshan:1963ts} (see \cite{Campo:2005sy,Campo:2008ju} for applications to cosmological perturbations). 
As noted in \cite{Campo:2005sy,Martin:2022kph}, a regular $P$-function representation also implies the loss of bipartite entanglement, so that the resulting state is separable.
Our interest is in determining the how the conditions for the regularity of the $P$-function translate into constraints on the decoherence parameters landscape.
For a Gaussian state, this occurs precisely when the fluctuations of the state are larger than the vacuum in all phase-space directions. 
Mathematically, this requires that the covariance matrix $\mathbf \Sigma_1$ minus the vacuum covariance matrix is positive semi-definite
\begin{equation}
    \Delta \Sigma = \mathbf{\Sigma}_1 - \mathbf{\Sigma}_{\rm vac} = \begin{pmatrix} \sigma_v^2 & \sigma_{vp} \\ \sigma_{vp} & \sigma_p^2 \end{pmatrix}- \begin{pmatrix} \frac{1}{2k} & 0 \\ 0 & \frac{k}{2} \end{pmatrix} \geq 0\, .
    \label{eq:PRepCondition}
\end{equation}
It is worth noting that this positive semi-definiteness condition is independent of the choice of partition between the two modes. 
For example, a change of partition from the $(\mathbf{k}, -\mathbf{k})$
Fourier modes to the real and imaginary standing-wave variables \eqref{eq:RealImaginaryVariables} corresponds to an orthogonal symplectic transformation $\mathbf{M}$ acting on the mode indices, under which 
$\Delta \Sigma \rightarrow \mathbf{M}\ \Delta \Sigma\ \mathbf{M}^T$. 
Since positive semi-definiteness 
is invariant under such orthogonal similarity transformations, the condition \eqref{eq:PRepCondition} for the regularity of the $P$-function is independent of the choice of partition between the two modes. 
A two-mode state with a regular $P$-function is thus also separable in every partition simultaneously.
This is in direct contrast to quantum discord and Bell inequality violations, which depend explicitly on the choice of subsystem partition \cite{Martin:2015qta}.

The positive semi-definite requirement \eqref{eq:PRepCondition} translates into two strict constraints on our landscape decoherence models.
First, the diagonal elements of the resulting matrix must be non-negative
\begin{equation}
    \sigma_v^2 \geq \frac{1}{2k},\qquad \sigma_p^2 \geq \frac{k}{2}\, .
\end{equation}
Because the amplitude is exponentially enhanced during inflation $\sigma_v^2 =e^{2r_k}/(2k)$, the first condition is trivially satisfied, while the second one indicates that the quantum-to-classical transition requires an \emph{enhanced} momentum variance, beyond the vacuum level.
Interestingly, this is in direct contrast to the common phenomenological approach towards classicalization 
in which the ``decaying mode" (corresponding here to the momentum variance) is set to zero.
Not only is the $\sigma_p^2 \rightarrow 0$ limit inaccessible for a finite amplitude variance, as discussed above, but in order to have the density matrix transition into a genuinely classical mixed state (with a regular, non-singular $P$-function), you must do the opposite: the environment must inject enough noise into the system to push the momentum variance beyond the vacuum limit $\sigma_p^2 \geq \sigma_{p,\rm vac}^2$.

The second condition for the matrix \eqref{eq:PRepCondition} to be positive definite
is that the determinant must be non-negative.
Expanding the determinant and expressing it in terms of the determinant of the covariance matrix \eqref{eq:WignerSubMatrix} yields
\begin{align}
    & \left(\sigma_v^2 - \frac{1}{2k}\right)\left(\sigma_p^2 - \frac{k}{2}\right) - \sigma_{vp}^2 \geq 0 \\
    \Rightarrow\ & \sigma_v^2 \sigma_p^2 - \sigma_{vp}^2 \geq \frac{k}{2} \sigma_v^2 + \frac{1}{4} \frac{\sigma_p^2}{\sigma_{p,\rm vac}^2} - \frac{1}{4}\, .
\end{align}
Using the relationship between the determinant and the purity \eqref{eq:PurityCovarianceDefn}, and taking the highly-squeezed limit $\sigma_v^2 = e^{2r_k}/(2k)$, we can re-cast this constraint in terms of our landscape parameters
\begin{align}
     \mu \leq \frac{1}{e^{2r_k} + \frac{\sigma_p^2}{\sigma_{p,\rm vac}^2} -1}\,.
     \label{eq:PFuncPurityBound}
\end{align}
This result demonstrates that ``classicality," in the sense of a state that can be represented as a classical stochastic distribution when quantum fluctuations are removed, places a strict upper bound on the purity required to achieve a fundamentally classical phase-space distribution. 
On our landscape, this carves out a subspace in which the $P$-function of the state can be rigorously interpreted as a classical probability, as seen in the lightly shaded region of Figure \ref{fig:LandscapePlotEmpty}.
Outside of this region the $P$-function exhibits highly singular quantum pathologies (e.g., infinite derivatives of Dirac delta functions).
Because the inflationary squeezing is exponentially large ($e^{2r_k} \gg 1$), the right-hand side of \eqref{eq:PFuncPurityBound} is typically dominated by the exponential amplitude, yielding an upper bound on the purity
\begin{equation}
\mu \lesssim e^{-2r_k}\, .
\label{eq:PFuncPurityBoundApprox}
\end{equation}
A similar constraint was found in \cite{Martin:2022kph} by demanding separability in the $(\mathbf k,-\mathbf k)$ partition.
Therefore, for the quantum state of cosmological perturbations to be interpreted as a purely classical stochastic distribution, the system must have been sufficiently decohered to suppress its purity to at least $\mathcal{O}(e^{-2r_k})$, acting as the horizontal ceiling for ``classicality" on our parameter space.

\subsection{Specific Models}
\label{subsec:specific_models}

\begin{figure}[t]
\centering\includegraphics[width=0.9\textwidth]{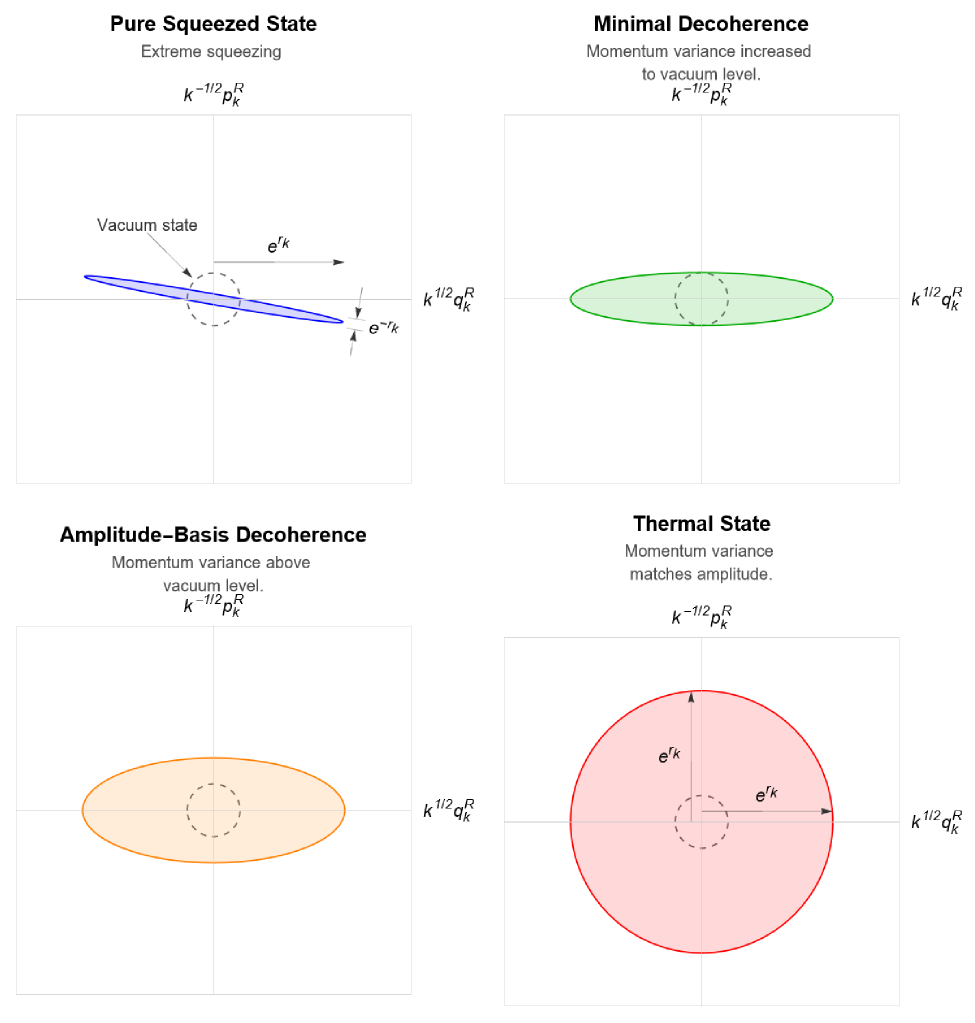}
\caption{The $1\sigma$ contours of the Gaussian Wigner function \eqref{eq:WignerMixed} in $(k^{1/2} q_k^R,k^{-1/2}p_k^R)$ phase space for several different states. The variance in the amplitude $q_k$ is fixed by the amplitude of observed curvature perturbations.
(Top Left) A pure squeezed state is a stretched rotated ellipse, with semi-minor axis exponentially suppressed compared to the variance of a vacuum state, but momentum variance equal to that of the vacuum.
(Top Right) The minimal decoherence model \eqref{eq:MinimalTwoModeCoherentDensity} reduces the quantum coherence just enough so that the variance of the semi-minor axis is equal to that of a vacuum state. 
(Bottom Left) Decoherence into a density matrix that is diagonal in amplitude basis leads to a variance much larger than the vacuum.
(Bottom Right) A Gaussian thermal density matrix is isotropic in phase space, with a variance in the momentum direction equal to that of the amplitude, exponentially larger than the vacuum.}
\label{fig:WignerModelPlots}
\end{figure}

Having established the strict mathematical and physical boundaries of the mixed-state landscape, we can now locate specific physical models of decoherence within this parameter space. 
To build physical intuition for these distinct decoherence mechanisms, Figure \ref{fig:WignerModelPlots} illustrates the $1\sigma$ phase-space contours of their corresponding Wigner functions, highlighting how different environmental interactions alter the fundamental vacuum fluctuations.
Table \ref{tab:decoherence_models} summarizes the dimensionless landscape parameters (the dimensionless momentum variance and purity) for each of the models discussed below, evaluated in the superhorizon, strong-squeezing limit ($r_k \gg 1$).
Finally, Figure \ref{fig:LandscapePlotModels} serves as the ultimate map of our parameter space, pinpointing the exact coordinates of each model on the landscape of mixed states of cosmological perturbations.

\subsubsection{Minimal Coherent State Decoherence}

Having defined the landscape of allowed mixed states, we can now locate specific physical models of decoherence in this parameter space. 
We begin by considering minimal deviations from the pure two-mode squeezed state
through the perspective of coherent states \cite{Matacz:1992tp,Campo:2004sz,Campo:2005sy}. 
In this formalism, the pure two-mode squeezed state $|\Psi_{\mathbf k}\rangle$ can be recast as a continuous, correlated superposition of coherent state pairs
\begin{equation}
|\Psi_{\mathbf{k}}\rangle = \int \frac{d^2\alpha}{\pi} \mathcal{A}_k(\alpha) |\alpha_{\mathbf{k}}\rangle \otimes |z_k \alpha^*_{-\mathbf{k}}\rangle\, ,
\end{equation}
where $\alpha_{\mathbf k}$ is a complex coherent amplitude, the quantity
\begin{equation}
\mathcal{A}_k(\alpha) = \frac{1}{\cosh r_k}\ \exp\left(-\frac{|\alpha_{\mathbf k}|^2}{2 \cosh^2 r_k}\right)
\end{equation}
is a Gaussian probability amplitude to find the mode $\mathbf k$ in the coherent state $|\alpha_{\mathbf k}\rangle$, and the correlation parameter $z_k = e^{2i\phi_k} \tanh r_k$ is dictated entirely by the standard squeezing parameters. 
Because the coherent states are overcomplete, the probability of finding the pure two-mode squeezed state $|\Psi_{\mathbf{k}}\rangle$ in a specific pair of individual one-mode coherent states, $|\alpha_{\mathbf{k}}\rangle \otimes |\beta_{-\mathbf{k}}\rangle$, is given by the squared projection 
\begin{equation}
P(\alpha, \beta) = |\langle \alpha_{\mathbf{k}}, \beta_{-\mathbf{k}} |\Psi_{\mathbf k}\rangle|^2 = |\mathcal{A}_k(\alpha)|^2 \exp\left(-|\beta - z_k \alpha^*|^2\right)\, .
\label{eq:2ModeCoherentJointProb}
\end{equation}
This joint probability distribution captures the bipartite entanglement of the system. If an observer were to measure the $\mathbf{k}$ mode and find it with a coherent amplitude $\alpha$, the conditional probability of finding its $-\mathbf{k}$ partner with an amplitude $\beta$ is a sharp Gaussian centered around $\beta = z_k \alpha^*$.
From this state, we can construct the pure state density matrix in the two-mode coherent basis. 
Taking the outer product $\hat{\rho}_{\mathbf{k}}^{\text{pure}} = |\Psi_{\mathbf{k}}\rangle\langle\Psi_{\mathbf{k}}|$ yields a double integral over the independent complex amplitudes $\alpha$ and $\alpha'$
\begin{align}
\hat{\rho}_{\mathbf{k}}^{\text{pure}} &= \int \frac{d^2\alpha}{\pi} \int \frac{d^2\alpha'}{\pi} \mathcal{A}_k(\alpha) \mathcal{A}_k^*(\alpha') \left( |\alpha_{\mathbf{k}}\rangle\langle \alpha'_{\mathbf{k}}| \otimes |z_k \alpha^*_{-\mathbf{k}}\rangle \langle z_k {\alpha'}^*_{-\mathbf{k}}| \right) \\
 &= \int \frac{d^2\alpha}{\pi} \int \frac{d^2\alpha'}{\pi} \frac{1}{\cosh^2 r_k} e^{-\frac{|\alpha|^2 + |\alpha'|^2}{2 \cosh^2 r_k}} \left( |\alpha_{\mathbf{k}}\rangle\langle \alpha'_{\mathbf{k}}| \otimes |z_k \alpha^*_{-\mathbf{k}}\rangle \langle z_k {\alpha'}^*_{-\mathbf{k}}| \right)\,.
 \label{eq:PureStateDensityCoherent}
\end{align}
In this pure state, the off-diagonal coherence ($\alpha \neq \alpha'$) is fully preserved and weighted by the Gaussian distributions.

We must now consider how this system behaves when coupled to an environment. Following the arguments of \cite{Zurek_Coherent_States_Decoherence,Campo:2004sz}, when a harmonic oscillator interacts with a generic environment, the coherent states naturally emerge as a ``pointer basis." 
According to this line of reasoning, coherent states are robust states against environmental monitoring because they minimize entropy production and correspond to highly localized, classical-like trajectories in phase space.
As a result of this environmental interaction, the density matrix rapidly loses its off-diagonal interference terms when expressed in the coherent state basis. 
When the off-diagonal macroscopic coherences are suppressed, we are left with a density matrix whose elements are simply given by the classical probabilities of finding the system in those specific coherent states.

Applying this logic to the cosmological pure state, let us consider that the environment continuously measures the system, forcing it into a statistical mixture of the two-mode coherent state pointer states. 
This motivates the so-called ``minimal decoherence" density matrix (the reason for this name will become clear below)
\begin{equation}
\hat{\rho}_{\text{min}} = \int \frac{d^2\alpha}{\pi} \int \frac{d^2\beta}{\pi} P(\alpha, \beta) \left( |\alpha_{\mathbf{k}}\rangle\langle\alpha_{\mathbf{k}}| \otimes |\beta_{-\mathbf{k}}\rangle\langle\beta_{-\mathbf{k}}| \right)\, ,
\label{eq:MinimalTwoModeCoherentDensity}
\end{equation}
where $P(\alpha,\beta)$ is given by the joint probability \eqref{eq:2ModeCoherentJointProb} to find the original pure squeezed state in the individual coherent states.
By replacing the pure state double-integral—which possessed exact phase coherence between arbitrary amplitudes $\alpha$ and $\alpha'$—with this diagonalized mixture, we transition from a quantum superposition of correlated pairs to a classical ensemble of coherent states.
Notice that $P(\alpha,\beta)$ is precisely the corresponding $P$-function for $\hat \rho_{\rm min}$, so this decoherence scheme is equivalent to demanding that the system has a regular, positive-definite $P$-function given by \eqref{eq:2ModeCoherentJointProb}.
Thus, this model is ``classical," in the sense discussed in Section \ref{subsec:ClassicalPFunction}.

\begin{table}[tbp]
    \centering
    \renewcommand{\arraystretch}{1.8}
    \begin{tabular}{@{}llcc@{}}
        \toprule
        \textbf{Model} & \textbf{Symbol} & \begin{tabular}[t]{@{}c@{}}\textbf{Momentum Variance} \\[-1ex] $\left(\sigma_p^2 / \sigma_{p,\text{vac}}^2\right)$\end{tabular} & \textbf{Purity} $(\mu)$ \\
        \midrule
        Pure Squeezed State & $\hat{\rho}_{\text{pure}}$ & $1$ & $1$ \\
        Minimal Decoherence & $\hat{\rho}_{\text{min}}$ & $3$ & $\frac{1}{2}e^{-2r_k}$ \\
        $\delta_k$ Decoherence & $\hat{\rho}_{\delta}$ & $1 + \delta_k$ & $\frac{1}{1 + \delta_k e^{2r_k}}$ \\
        Amplitude-Diagonal & $\hat{\rho}_{\text{diag}}$ & $\sim \frac{1}{288}(\epsilon+\eta)^2\Delta_\zeta^2 e^{r_k}$ & $\sim \frac{288 e^{-3r_k}}{(\epsilon+\eta)^2\Delta_\zeta^2}$ \\
        Phase-Averaged Thermal & $\hat{\rho}_{\text{th}}$ & $e^{2r_k}$ & $2e^{-2r_k}$ \\
        Uncorrelated Thermal & $\hat{\rho}_{\text{uc}}$ & $e^{2r_k}$ & $e^{-4r_k}$ \\
        \bottomrule
    \end{tabular}
    \caption{Summary of the dimensionless landscape parameters of various decoherence models from Section \ref{sec:decoherence} on the mixed-state landscape. The dimensionless momentum variance and purity are evaluated in the superhorizon, strong-squeezing limit ($r_k \gg 1$).}
    \label{tab:decoherence_models}
\end{table}

\begin{figure}[t]
\centering\includegraphics[width=.9\textwidth]{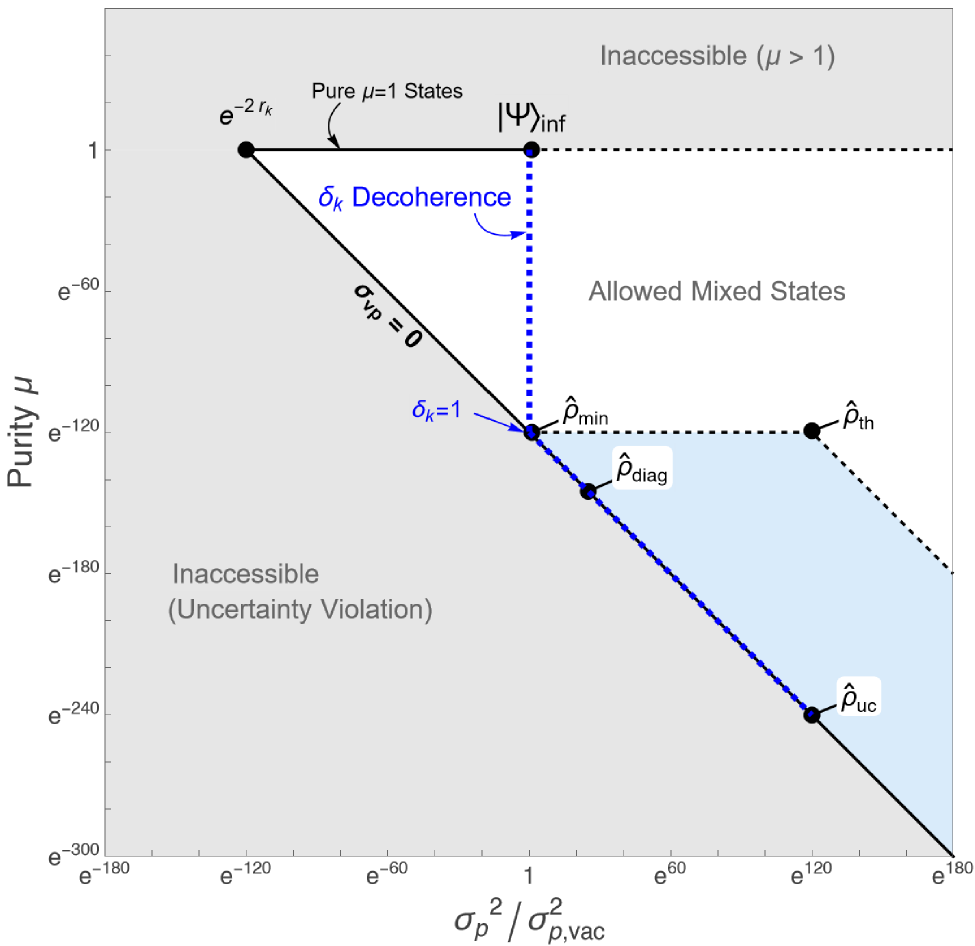}
\caption{The location of several proposed models of decoherence of inflationary perturbations on the decoherence landscape discussed in this section, all of which are on the boundary or inside of the P-representable (blue shaded) region (see Figure \ref{fig:LandscapePlotEmpty}). 
$\hat \rho_{\rm min}$: ``Minimal decoherence," diagonal in the basis of coherent states \eqref{eq:MinimalTwoModeCoherentDensity};
(dashed blue) $\delta_k$: The one-parameter family of decoherence models \eqref{eq:deltaCoherenceDefinition} which interpolates between the pure two-mode squeezed state as $\delta_k \rightarrow 0$, the minimal decoherence model \eqref{eq:MinimalTwoModeCoherentDensity} for $\delta_k = 1$, and the uncorrelated thermal state \eqref{eq:IndependThermalDensity} as $\delta_k\rightarrow \sinh^2 r_k$;
$\hat \rho_{\text{diag}}$: A density matrix that is diagonal in amplitude basis, \eqref{eq:diagDensityMatrixZeta};
$\hat \rho_{\rm th}$: The phase-averaged two-mode thermal state \eqref{eq:2ModeThermal}; $\hat \rho_{\rm uc}$: An uncorrelated thermal ``classical" state \eqref{eq:ClassicalDensity}.  }
\label{fig:LandscapePlotModels}
\end{figure}

To see the consequences of this minimal decoherence scheme, we evaluate the joint probability $P(\alpha, \beta)$ in the strong squeezing limit relevant for cosmological scales ($r_k \gg 1$). 
In this limit, the Gaussian distribution is extremely broad with respect to the $\alpha$ variable, but comparatively narrow with respect to $\beta$. 
Consequently, the distribution effectively acts as a Dirac delta function centered strictly on the classical correlation, $P(\alpha, \beta) \propto \delta^2(\beta - z_k \alpha^*)$.
This drastically simplifies the minimal decoherence density matrix, collapsing the double integral into a single integral over $\alpha$:
\begin{equation}
\hat{\rho}_{\text{min}} \approx \int \frac{d^2\alpha}{\pi} \frac{1}{\cosh^2 r_k} e^{-|\alpha|^2 / \cosh^2 r_k} \left( |\alpha_{\mathbf{k}}\rangle\langle\alpha_{\mathbf{k}}| \otimes |z_k \alpha^*_{-\mathbf{k}}\rangle\langle z_k \alpha^*_{-\mathbf{k}}| \right)\,.
\end{equation}
Comparing this directly to the pure-state density matrix \eqref{eq:PureStateDensityCoherent}, it is clear that the macroscopic off-diagonal coherence has been entirely suppressed, leaving a diagonal mixture of perfectly correlated coherent states.

Because this diagonalized mixture remains a Gaussian state, we can analytically calculate the two-point functions of the Hermitian phase-space (Mukhanov) variables for this minimal decoherence scheme
\begin{align}
\sigma_{v,\text{min}}^2 &= \frac{1}{2k} \left[ \cosh(2r_k) - \cos(2\phi_k)\sinh(2r_k) + 2 \right]\,; \\
\sigma_{p,\text{min}}^2 &= \frac{k}{2} \left[ \cosh(2r_k) + \cos(2\phi_k)\sinh(2r_k) + 2 \right]\,; \\
\sigma_{vp,\text{min}} &= \frac{1}{2} \sinh(2r_k)\sin(2\phi_k)\, .
\end{align}
Notice the striking mathematical simplicity of this result: the effect of environment-induced decoherence in the coherent state basis is equivalent to simply adding a constant value of $2$ to the bracketed terms of the amplitude and momentum variances, while leaving the cross-correlation entirely unaffected.

To understand the physical implications of this, we evaluate these variances in the inflationary limit \eqref{eq:dSSolution},\eqref{eq:dSSolution:2} ($r_k \gg 1, \phi_k \approx \pi/2$).
The amplitude variance remains dominated by the exponential squeezing, $\sigma_{v,\text{min}}^2 \approx e^{2r_k}/(2k)$, and the covariance evaluates to $\sigma_{vp,\text{min}} \approx \frac{1}{2} e^{r_k}$. 
However, the crucial deviation occurs in the momentum variance, increasing it from its vacuum value
\begin{equation}
\sigma_{p,\text{min}}^2 \approx \frac{3k}{2} = \sigma_{p,\text{vac}}^2 + k = 3 \sigma_{p,\text{vac}}^2 \, .
\end{equation}
Because it preserves the exact pure-state behavior of the enhanced amplitude and covariance, altering the momentum variance by only a minimal $\mathcal{O}(k)$ addition, this framework is naturally referred to as ``minimal decoherence," illustrated graphically in Figure \ref{fig:WignerModelPlots}.
While changing the momentum variance by one unit of $k$ seems like a relatively minor alteration compared to the $e^{2r_k}$ scales in the amplitude, the state is no longer pure and has vanishingly small purity
\begin{equation}
\mu_{\text{min}} = \frac{1}{4 \det \mathbf{\Sigma}_1} = \frac{1}{4(\sigma_{v,\text{min}}^2\sigma_{p,\text{min}}^2 - \sigma_{vp,\text{min}}^2)} \approx \frac{1}{2} e^{-2r_k}\,,
\end{equation}
placing this state near the lower boundary of the landscape of decoherence models in Figure \ref{fig:LandscapePlotModels}.

\subsubsection{One-Parameter Family of Decoherence: \texorpdfstring{$\delta_k$}{delta\_k}}
\label{subsubsec:deltaDecoherence}

Motivated by the minimal decoherence scheme, which demonstrated how environment-induced corrections suppress coherence while leaving macroscopic amplitudes largely untouched, we now consider a more general deformation of the quantum state, following \cite{Campo:2004sz}. 
To systematically explore this, it is highly advantageous to parameterize the state directly in terms of the expectation values of the creation and annihilation operators.
Assuming spatial isotropy and momentum conservation, the two-point functions of the standard ladder operators are fully characterized by the occupation number $n_k$ and the (complex) quantum coherence $\mathcal{C}_k$, defined respectively as
\begin{align}
\label{eq:deltaOccupation}
\langle \hat{c}_{\mathbf{k}}^\dagger \hat{c}_{\mathbf{k}'} \rangle &= n_k\ (2\pi)^3 \delta^3(\mathbf{k} - \mathbf{k}')\,; \\
\label{eq:deltaCoherence}
\langle \hat{c}_{\mathbf{k}} \hat{c}_{-\mathbf{k}'} \rangle &= \mathcal{C}_k\ (2\pi)^3 \delta^3(\mathbf{k} + \mathbf{k}')\,;
\end{align}
generalizing the pure state two-point functions \eqref{eq:PureOccupation},\eqref{eq:PureCoherence}.
For the pure two-mode squeezed state generated by standard unitary inflationary evolution, these quantities are strictly dictated by the squeezing parameters
\begin{align}
n_k &= \sinh^2(r_k)\,; \\
\mathcal{C}_k &= -\frac{1}{2} e^{-2i\phi_k} \sinh(2r_k)\, .
\end{align}
By expressing the coherence in terms of the occupation number, we find the exact identity for the pure state: $|\mathcal{C}_k|^2 = n_k (n_k + 1)$. This relation represents the absolute maximum amount of quantum coherence a state can possess for a given energy (occupation number).

We can phenomenologically model decoherence by considering a one-parameter family of deformations from this pure state. 
We assume an environmental interaction that preserves the particle number $n_k$ but degrades the phase correlations between the opposite momentum modes. 
We parameterize this loss of coherence by introducing a real deformation parameter $\delta_k$, such that
\begin{equation}
|\mathcal{C}_k|^2 = n_k (n_k + 1 - \delta_k)\,.
\label{eq:deltaCoherenceDefinition}
\end{equation}
To ensure the coherence remains physically valid ($|\mathcal{C}_k|^2 \ge 0$), the deformation parameter is bounded within the range $0 \le \delta_k \le n_k + 1 = \cosh^2 r_k$. 
The lower bound ($\delta_k = 0$) perfectly recovers the pure two-mode squeezed state. 
Conversely, the upper bound ($\delta_k = n_k + 1$) completely destroys the coherence ($|\mathcal{C}_k| = 0$), recovering the uncorrelated product of one-mode thermal states \eqref{eq:IndependThermalDensity} discussed in Section \ref{subsubsec:Thermal}.

In addition to interpolating between the pure two-mode squeezed state and the uncorrelated thermal state, the $\delta_k$ parameterization leads to a direct connection to the stochastic classical description of the state via the Glauber-Sudarshan $P$-function discussed earlier. 
Requiring the state described by \eqref{eq:deltaCoherenceDefinition} to be described by a regular, non-negative P-function 
\eqref{eq:PRepCondition} yields a strict threshold for the decoherence parameter in these variables \cite{Campo:2005sy}
\begin{equation}
n_k^2 \ge |\mathcal{C}_k|^2 \implies 
\delta_k \ge 1\, .
\end{equation}
Thus, as the environment degrades the coherence, the system crosses a fundamental threshold at $\delta_k = 1$. 
For any deformation $\delta_k \ge 1$, the cosmological state can be represented by a classical stochastic distribution. 
This makes $\delta_k$ a useful parameterization to vary the amount of decoherence to interpolate from a pure state to a state that can be described as a classical stochastic distribution.

We can directly calculate the variances of $\hat v_{\mathbf k},\hat p_{\mathbf k}$ from \eqref{eq:CreationAnnihilationDefn} and \eqref{eq:deltaOccupation},\eqref{eq:deltaCoherence} 
assuming that the environmental interaction preserves the squeezing angle-dependence of the coherence, e.g. ${\mathcal C}_k = -e^{-2i\phi_k}|{\mathcal C}_k|$
\begin{align}
    \sigma_v^2 &= \frac{1}{2k} \left[\cosh(2r_k) - \cos(2\phi_k)\sinh(2r_k) \sqrt{1-\frac{\delta_k}{\cosh^2 r_k}}\ \right]\,; \\
    \sigma_p^2 &= \frac{k}{2} \left[\cosh(2r_k) + \cos(2\phi_k)\sinh(2r_k) \sqrt{1-\frac{\delta_k}{\cosh^2 r_k}}\ \right]\,; \\
    \sigma_{vp} &= \frac{1}{2} \sin(2\phi_k) \sinh(2r_k) \sqrt{1-\frac{\delta_k}{\cosh^2 r_k}}\,.
\end{align}
For $\delta_k\neq 0$, in the inflationary limit 
the variances and covariance become
\begin{align}
    \sigma_v^2 &\approx \frac{e^{2r_k}}{2k}\,; \\
    \sigma_p^2 &\approx \frac{k}{2} \left(1+\delta_k\right) = \sigma_{p,\text{vac}}^2 (1+\delta_k)\,; \\
    \sigma_{vp} &\approx \frac{1}{2}e^{r_k}\,.
\end{align}
This result is highly illuminating. 
The amplitude variance $\sigma_v^2$ remains dominated by the $e^{2r_k}$ term, ensuring that this entire family of states naturally remains consistent with the macroscopic scale of CMB observations.
However, the momentum variance is additively increased above its vacuum value by the decoherence parameter $\delta_k$, similarly to the minimal decoherence model of the previous subsection.
As we noticed there, even though this is a relatively minor change in the momentum variance, the purity is strongly dependent on $\delta_k$
\begin{equation}
    \mu_{\delta} \approx \frac{1}{1+\delta_k e^{2r_k}}\, .
\end{equation}
This family of decoherence models, with $0 \leq \delta_k \leq \cosh^2 r_k$, parameterizes a line on the landscape of decoherence models, as seen in Figure \ref{fig:LandscapePlotModels}.
Starting at $\delta_k = 0$, the state becomes the pure two-mode squeezed state with $\mu_\delta = 1$. 
As the environment induces decoherence and $\delta_k$ increases, the purity sharply decreases to $\mu_\delta \approx \delta_k e^{-2r_k}$, quickly approaching the lower bound of the allowed landscape of mixed states.

This $\delta_k$-decoherence parameterization provides a kinematic baseline for classicality. 
The minimal decoherence model (which added roughly $\sigma_{p,\text{vac}}^2$ to the momentum variance) pushes the state exactly across this threshold. 
We now examine how specific dynamical interactions drive the state further into this classical regime.

\subsubsection{Diagonal in Amplitude Basis}

While the minimal decoherence and $\delta_k$ models provide a useful kinematic parameterization,
a more dynamically rigorous approach to decoherence involves treating the cosmological perturbations as an open quantum system interacting with an environment, such as short-wavelength sub-Hubble modes or other spectator fields \cite{Martineau:2006ki, Burgess:2006jn, Burgess:2014eoa, Shandera:2017qkg}.
In this framework, continuous monitoring by the environment dynamically selects a preferred ``pointer basis"; 
detailed analysis suggests that the enormous squeezing of the state naturally singles out the field configuration (amplitude) basis as the robust pointer basis \cite{Kiefer:2006je, Kiefer:2008ku, Burgess:2014eoa, Nelson:2016kjm, Burgess:2022nwu}.

When the environment is traced out, the resulting reduced density matrix $\rho_{\text{diag}}(\zeta_{\mathbf k}, \tilde \zeta_{\mathbf k})$ for the comoving curvature perturbation $\zeta_{\mathbf k}$ becomes sharply peaked along the diagonal. 
Phenomenologically, the off-diagonal terms ($\zeta \neq \zeta'$) are exponentially suppressed by a decoherence factor \cite{Nelson:2016kjm}
\begin{equation}
\rho_{\text{diag}}(\zeta_{\mathbf{k}}, \tilde\zeta_{\mathbf k}) \sim \rho_{\text{pure}}(\zeta_{\mathbf k},\tilde \zeta_{\mathbf k})\times\exp\left[-\frac{1}{2}\Gamma_{\text{decoh}} \frac{|\zeta_{\mathbf k} - \tilde \zeta_{\mathbf k}|^2}{\Delta_{\zeta}^2}\right]\,,
\label{eq:diagDensityMatrixZeta}
\end{equation}
where $\rho_{\rm pure}$ is the density matrix for a pure state (see e.g.~\eqref{eq:RIDensityMatrix}), $\Gamma_{\text{decoh}}$ represents the decoherence rate, and $\Delta_\zeta^2$ is the amplitude of curvature perturbations.
Rigorous derivations using Lindblad equations and effective field theory approaches demonstrate that tracing out sub-Hubble environment modes generically produces a decoherence rate that scales with the physical volume, yielding a $(aH/k)^3$ dependence \cite{Burgess:2014eoa,Nelson:2016kjm}. 
Specifically, considering decoherence sourced by minimal gravitational nonlinearities, the rate takes the  form $\Gamma_{\text{decoh}} \propto (\epsilon+\eta)^2 \Delta_\zeta^2 \left(\frac{aH}{k}\right)^3$, where $\epsilon,|\eta|\ll1$ are the inflationary slow-roll parameters.
Decoherence occurs for $\Gamma_{\text{decoh}}\gtrsim {\mathcal O}(1)$, leading to a density matrix that is strongly peaked along the diagonal elements.

Translating \eqref{eq:diagDensityMatrixZeta} into the Mukhanov coordinates $v_{\mathbf{k}} = z(\eta) \zeta_{\mathbf{k}}$ leads to a density matrix of the form
\begin{equation}
    \rho_{\text{diag}}(v_{\mathbf k},\tilde v_{\mathbf k}) \approx 
    \rho_{\rm pure}(v_{\mathbf k},\tilde v_{\mathbf k})\times \exp\left[-\frac{k}{288} (\epsilon+\eta)^2\Delta_{\zeta}^2e^{r_k}\ |v_{\mathbf k}-\tilde v_{\mathbf k}|^2\right]\,,
    \label{eq:DiagAmplitudeDensity}
\end{equation}
where we used $(aH/k) = e^{r_k}$ as the amount of squeezing since the mode exited the horizon. This density matrix is Gaussian, 
with corresponding
\begin{equation}
    \sigma_{p,\text{diag}}^2 = \frac{k}{2} + \frac{k}{144}(\epsilon+\eta)^2\Delta_{\zeta}^2e^{r_k}\,.
\end{equation}
Evaluating this for slow-roll inflationary parameters $\epsilon,|\eta| \sim 10^{-2}$ and using $\Delta_{\zeta}^2 = 2\times 10^{-9}$, the dynamic vanishing of the off-diagonal coherence occurs at an intermediate value of the momentum variance 
\begin{equation}
    \frac{\sigma_{p,\text{diag}}^2}{\sigma_{p,\text{vac}}^2} = 1+ \frac{1}{288}(\epsilon+\eta)^2\Delta_{\zeta}^2\ e^{r_k} \approx e^{25} 
    \label{eq:DiagonalMomentumVariance}
\end{equation}
for modes experiencing $r_k\sim 60$ e-folds of squeezing.
This is exponentially larger than the vacuum, as illustrated in Figure \ref{fig:WignerModelPlots}.
The corresponding purity
\begin{equation}
    \mu_{\text{diag}} \approx \frac{1}{4\sigma_v^2 \sigma_p^2 - \sigma_{vp}^2} \approx \frac{288 e^{-3r_k}}{(\epsilon+\eta)^2 \Delta_\zeta^2} \approx e^{-145} \ll 1\,,
\end{equation}
places this state on the lower boundary of the landscape of decoherence models in Figure \ref{fig:LandscapePlotModels}.

\subsubsection{Two-Mode Thermal State and Classical-Only Correlations}
\label{subsubsec:Thermal}

Finally, we consider the phenomenological extreme of decoherence: a complete loss of phase information.
If the environment interacts with the cosmological perturbations in a way that randomizes the squeezing angle $\phi_k$, we can model the resulting state by taking the pure two-mode squeezed density matrix $\hat{\rho}_{\mathbf{k}} = |\Psi_{\mathbf{k}}\rangle\langle\Psi_{\mathbf{k}}|$ as in \eqref{eq:2ModeSqueezedStateDensity} and averaging over $\phi_k \in [0, 2\pi]$ \cite{Brandenberger:1992sr,Brandenberger:1992jh,Prokopec:1992ia,Gasperini:1992xv,Gasperini:1993mq,Brahma:2020zpk}. 
Because the off-diagonal terms in the Fock basis carry phase dependence like $e^{i(n-m)\phi_k}$, this averaging process entirely erases the off-diagonal coherences. 
The resulting density matrix is diagonal in the number basis:
\begin{equation}
\hat \rho_{\rm th} = \frac{1}{2\pi} \int_0^{2\pi} d\phi_k\ |\Psi_{\mathbf{k}}\rangle\langle\Psi_{\mathbf{k}}| = \frac{1}{\cosh^2 r_k} \sum_{n=0}^\infty \tanh^{2n} (r_k) |n_{\mathbf{k}},n_{-\mathbf{k}}\rangle \langle n_{\mathbf{k}},n_{-\mathbf{k}}|\,.
\label{eq:2ModeThermal}
\end{equation}
This takes the form of a two-mode thermal state. Note that while this state is constructed by averaging Gaussian pure states, the resulting thermal mixture itself is non-Gaussian, as we will discuss below (it possesses a non-trivial four-point function). 
Consequently, it does not strictly possess a Gaussian covariance matrix. 
However, we can still evaluate its two-point functions directly from the density matrix to find its effective location in our parameter space.

The phase averaging destroys the cross-correlation, yielding $\sigma_{vp} = 0$, while the variance of the momentum becomes strictly tied to the variance of the amplitude
\begin{align}
    \langle \hat v_{\mathbf k} \hat v_{\mathbf k'}\rangle_{\rm th} &= \frac{1}{2k} \cosh(2r_k)\ (2\pi)^3 \delta^3(\mathbf{k}+\mathbf{k}')\,; \\ 
    \langle \hat p_{\mathbf k} \hat p_{\mathbf k'}\rangle_{\rm th} &= k^2 \langle \hat v_{\mathbf k} \hat v_{\mathbf k'}\rangle_{\rm th} = \frac{k}{2} \cosh(2r_k)\ (2\pi)^3 \delta^3(\mathbf{k}+\mathbf{k}')\,,
\end{align}
implying $\sigma_p^2 = k^2 \sigma_v^2$.
The momentum variance is exponentially enhanced, $\sigma_p^2 \approx k e^{2r_k}/2$, rather than being equal to its vacuum value as in the pure state, as demonstrated in Figure \ref{fig:WignerModelPlots}.
Because of the non-Gaussian nature of the state \eqref{eq:2ModeThermal} some care is needed in calculating  its corresponding purity.
The exact purity can be written as
\begin{equation}
    \mu_{\rm th} = \text{Tr}(\hat \rho_{\rm th}^2) = \frac{1}{\cosh^4 r_k}\sum_{n=0}^\infty \tanh^{2n} r_k = \frac{1}{\cosh(2r)} \approx 2 e^{-2r_k}\, .
\end{equation}
This is significantly larger than what we would have obtained for the purity if we had used the derivation \eqref{eq:PurityCovarianceDefn} based on the covariance matrix, $\mu \sim e^{-4r_k}$. This difference is due to the effects of the non-Gaussian correlations, enhancing the true purity of the state \eqref{eq:2ModeThermal}.
The effects of decoherence can also be seen through the associated von-Neumann entropy
\begin{align}
    S_{\text{th}} &= - \text{Tr}\left(\hat \rho_{\rm th} \ln \hat \rho_{\rm th}\right) = (1+\sinh^2 r_k)\ln (1+\sinh^2 r_k) - \sinh^2 r_k \ln (\sinh^2 r_k) \\
    &\approx 2 r_k\, ,
\end{align}
in the limit of large squeezing $r_k \gg 1$.
This phase-averaged, two-mode thermal state is located in the interior of the space of mixed states on the landscape of decoherence models for cosmological perturbations, as shown in Figure \ref{fig:LandscapePlotModels}.

As noted previously, the phase-averaged two-mode thermal state is fundamentally non-Gaussian.
This departure from Gaussianity is explicitly captured by the connected part of its four-point correlation function (the trispectrum) (see also \cite{Martin:2015qta}) 
\begin{align}
    \langle\hat v_{\mathbf k_1}\hat v_{\mathbf k_2}\hat v_{\mathbf k_3}\hat v_{\mathbf k_4}\rangle_c = \frac{\sinh^2(2 r_k)}{4k_1^2} (2\pi)^3\delta^3(\sum_i \mathbf k_i)\left[(2\pi)^6\delta^3(\mathbf k_1 + \mathbf k_2)\delta^3(\mathbf k_1 - \mathbf k_3) + \text{perms}\right]\, ,
\label{eq:ThermalTrispectrumV}
\end{align}
where the permutations denote additional symmetric momentum exchange channels (specifically $\mathbf k_1 = \mathbf k_2 = -\mathbf k_3 = -\mathbf k_4$ and $\mathbf{k}_1 = \mathbf{k}_4 = -\mathbf{k}_2 = -\mathbf{k}_3$).
Translating this into the curvature perturbation variable $\zeta_{\mathbf k}$ and evaluating in the strong squeezing limit ($r_k \gg 1$) during inflation, the curvature perturbation trispectrum takes the form
\begin{align}
    \langle\hat \zeta_{\mathbf k_1}\hat \zeta_{\mathbf k_2}\hat \zeta_{\mathbf k_3}\hat \zeta_{\mathbf k_4}\rangle_c = \frac{1}{16} \left(\frac{H_{dS}^2}{M_p^2 \epsilon}\right)^2 \frac{(2\pi)^3\delta^3(\sum_i \mathbf k_i)}{k_1^6}\left[(2\pi)^6\delta^3(\mathbf k_1 + \mathbf k_2)\delta^3(\mathbf k_1 - \mathbf k_3) + \text{perms}\right]\, .
\label{eq:ThermalTrispectrum}
\end{align}
Interestingly, on superhorizon scales this four-point function approaches a constant amplitude by the square of the two-point scalar amplitude $\Delta_\zeta^2 \sim H_{dS}^2/(M_p^2\epsilon)$.
However, its unique structure enforced by the product of multiple Dirac delta functions does not cleanly map onto standard primordial trispectrum templates (such as the local, equilateral, or orthogonal shapes studied in \cite{Planck:2019kim})
As a result, current observational constraints on non-Gaussianity cannot be directly applied to bound this specific two-mode thermal state model.

Importantly, this relationship between the amplitude and momentum variances is not unique to the thermal state; it is a universal feature of any state possessing strictly classical correlations between the $\mathbf{k}$ and $-\mathbf{k}$ sectors. 
To see this, we can formulate the problem using quantum discord. 
A bipartite state with zero quantum discord contains only classical correlations and can be written as a separable statistical mixture of local states. 
Assuming spatial isotropy, the most general density matrix of this form in the original $(\mathbf k,-\mathbf k)$ basis is
\begin{equation}
\hat \rho_{\rm sep} = \sum_{n,m} p_{nm} |n_{\mathbf{k}}\rangle \langle n_{\mathbf{k}}| \otimes |m_{-\mathbf{k}}\rangle\langle m_{-\mathbf{k}}|\,,
\label{eq:ClassicalDensity}
\end{equation}
where $p_{nm} = p_{mn}$ is a symmetric classical probability distribution.
Because this state is strictly diagonal in the local Fock bases, the expectation values of operator pairs that create or destroy particles in tandem, such as $\langle \hat{c}_{\mathbf{k}} \hat{c}_{-\mathbf{k}} \rangle$ and $\langle \hat{c}_{\mathbf{k}}^\dagger \hat{c}_{-\mathbf{k}}^\dagger \rangle$, are identically zero.
As with the phase-averaged thermal state above, the cross-correlations between the amplitude and momentum vanish, while the two-point functions of the amplitude and momentum are again locked together
\begin{align}
    \langle \hat v_{\mathbf k} \hat v_{\mathbf k'}\rangle_{\rm sep} &= \frac{1}{2k} \left( 1 + \sum_{n,m} 2n p_{nm} \right)\ (2\pi)^3 \delta^3(\mathbf{k}+\mathbf{k}')\,;  \\
    \langle \hat p_{\mathbf k} \hat p_{\mathbf k'}\rangle_{\rm sep} &= k^2 \langle \hat v_{\mathbf k} \hat v_{\mathbf k'}\rangle_{\rm sep} = \frac{k}{2} \left( 1 + \sum_{n,m} 2n p_{nm} \right)\ (2\pi)^3 \delta^3(\mathbf{k}+\mathbf{k}')\, .
\end{align}
Thus, for any separable state with classical-only correlations, the momentum variance is strictly determined by the amplitude variance. Fixing the amplitude to the CMB $1 + \sum_{n,m} 2n p_{nm} = \sigma_v^2 \sim e^{2r_k}$, the momentum variance is again enhanced compared to the pure-state $\sigma_p^2 = k^2 \sigma_v^2 \approx ke^{2r_k}/2$.

As with the phase-averaged two-mode thermal state, in order to place this  state on the landscape of cosmological perturbation models we need to calculate the exact purity
\begin{equation}
    \mu_{\rm sep} = \text{Tr}(\hat \rho_{\rm sep}^2) = \sum_{n,m} p_{nm}^2\,.
\end{equation}
Without specifying the distribution $p_{nm}$ further, we are not able to calculate the purity in general. As we saw before, if we assume the $(\mathbf k,-\mathbf k)$ modes to be correlated in particle number so that $p_{nm} = p_n \delta_{nm}$ with $p_n = \tanh^2(r_k)$ given by a thermal distribution, we are led to a purity $\mu \sim e^{-2r_k}/2$.
Alternatively, if we assume the modes are completely uncorrelated, statistically independent thermal states $p_{nm} = p_n p_m$ then the density matrix becomes
\begin{equation}
    \hat\rho_{\rm uc} = \left(\sum_n p_n |n_{\mathbf k}\rangle \langle n_{\mathbf k}|\right) \otimes \left(\sum_m p_m |m_{-\mathbf k}\rangle \langle m_{-\mathbf k}|\right)\,,
    \label{eq:IndependThermalDensity}
\end{equation}
and the purity is reduced further to $\mu_{\rm uc} \sim 1/\cosh^2(2r_k) \sim e^{-4r_k}\sim e^{-240}$.
This state is Gaussian, with momentum variance $\sigma_p^2/\sigma_{p,\rm vac}^2 = e^{2r_k} \sim e^{120}$, and represents an isotropic distribution in phase space, as shown in Figure \ref{fig:WignerModelPlots}.
This state is located at the extreme bottom-right of the parameter space of cosmological perturbations in Figure \ref{fig:LandscapePlotModels}.
In contrast to the phase-averaged two mode thermal state \eqref{eq:2ModeThermal}, this state is Gaussian, since the density matrix \eqref{eq:IndependThermalDensity} can be written as the product of two single-mode Gaussian thermal density matrices. 

Finally, we note a remarkable connection between this phenomenological thermal limit and the kinematic $\delta_k$ parameterization introduced in Section \ref{subsubsec:deltaDecoherence}. 
While setting $\delta_k = 1$ corresponds to the minimal decoherence necessary to cross the classicality threshold, the $\delta_k$ model is continuous and can be extended to represent more severe environmental degradation. 
If the environment interacts so aggressively that it drives the decoherence parameter to the extreme limit of $\delta_k \rightarrow \sinh^2 r_k \approx \frac{1}{4}e^{2r_k}$, the dimensionless momentum variance grows to $\sigma_p^2 / \sigma_{p,\text{vac}}^2 \sim e^{2r_k}$ and the purity drops to $\mu \sim e^{-4r_k}$, which are precisely the coordinates of the uncorrelated thermal state $\hat{\rho}_{\text{uc}}$. 
Therefore, the $\delta_k$ parameterization elegantly interpolates across the entire accessible classical landscape: its lower bound ($\delta_k = 1$) defines the minimal onset of classical stochasticity, while its upper limit ($\delta_k \rightarrow \sinh^2 r_k$) perfectly recovers the complete destruction of all bipartite correlations characteristic of the classical thermal state.

\section{Decoherence and the Decaying Mode}
\label{sec:decaying_mode}

In the previous section, we reviewed several proposals that aim to explain the decoherence and quantum-to-classical transition of cosmological perturbations.
Assuming the cosmological perturbations are Gaussian\footnote{With the notable exception of the two-mode phase-averaged thermal state of Section \ref{subsubsec:Thermal}.}, adiabatic, and posses a scalar amplitude matching that observed in the CMB, we mapped these models onto a landscape parameterized by the purity $\mu = \text{Tr}(\hat \rho^2)$ and the normalized momentum variance $\sigma_p^2/\sigma_{p,\text{vac}}^2$.

We now explore the phenomenological and observational consequences of these parameters. 
Assuming that the effects of environment-induced decoherence persist up to the end of inflation, these models establish specific boundary values for the
curvature perturbation $\zeta_{\mathbf k,\rm inf}(\eta_i)$ and its conformal time derivative $\zeta'_{\mathbf k, \rm inf}(\eta_i)$ at the transition out of the inflationary era ($\eta = \eta_i$). 

To rigorously translate the decohered quantum state into these classical boundary conditions, we must formalize the quantum-to-classical transition. 
As established in Section \ref{sec:decoherence}, environmental interactions suppress the off-diagonal interference terms of the density matrix.
Once this coherence is sufficiently degraded -- such as when crossing the classicality threshold discussed in Section \ref{subsubsec:deltaDecoherence} -- the quantum Wigner function becomes strictly positive-definite over the phase space. 
Mathematically, this allows us to reinterpret the quantum state not as a coherent superposition, but as a classical stochastic ensemble of distinct macroscopic field configurations. 
In this limit, quantum expectation values of Hermitian operators become formally indistinguishable from classical statistical ensemble averages.

Consequently, we can directly map the decohered quantum phase-space variances into classical stochastic power spectra.
Recall that the gauge-invariant curvature perturbation and its derivative are directly proportional to the Mukhanov variable and its conjugate momentum \eqref{eq:MukhanovMomentum}
\begin{equation}
\hat{v}_{\mathbf k} = z \hat{\zeta}_{\mathbf k}, \qquad \hat{p}_{\mathbf k} = z \hat{\zeta}'_{\mathbf k}\,,
\end{equation}
where $z = a M_{pl} \sqrt{2\epsilon}$.
Because all the decoherence models considered here preserve the macroscopic amplitude of the perturbations, 
the typical root-mean-square (RMS) amplitude of the curvature perturbation is unchanged by decoherence and matches the standard dimensionless scale observed in the CMB
\begin{equation}
\zeta_{\mathbf k,\rm inf}^{\rm rms}(\eta_i) \sim \sqrt{\Delta_\zeta^2(k)} \sim 10^{-5}\, .
\end{equation}

Conversely, the decoherence models differ drastically in their predictions for the momentum variance $\sigma_p^2$. A state with an enhanced momentum variance translates directly into a macroscopic, non-zero conformal time derivative $\zeta'_{\mathbf k}$. 
The power spectrum of this time-derivative is defined as
\begin{equation}
\langle \hat\zeta_{\mathbf k}' \hat \zeta_{\mathbf k'}'\rangle \equiv P_{\zeta'}(k)\ (2\pi)^3\ \delta^3(\mathbf k + \mathbf k')\, .
\label{eq:zetaPrimePowerSpectrum1}
\end{equation}
Using the relationship of $\zeta'_{\mathbf k}$ to the Mukhanov momentum $\hat p_{\mathbf k}$, this power spectrum is entirely determined by the momentum variance of the specific decoherence model
\begin{equation}
    \langle \hat \zeta_{\mathbf k}'\hat \zeta_{\mathbf k'}'\rangle  = \frac{\langle \hat p_{\mathbf k}\hat p_{\mathbf k'}\rangle}{z^2} \quad \Rightarrow \quad P_{\zeta'}(k) = \frac{\sigma_p^2}{z^2} = \frac{\sigma_p^2 H_{dS}^2 \eta_i^2}{2M_p^2 \epsilon}\,.
    \label{eq:zetaPrimePowerSpectrum2}
\end{equation}
To quantify the physical size of this derivative, 
we write its real-space variance via a logarithmic integral of the power spectrum 
\begin{equation}
\langle \zeta'(\mathbf x)^2\rangle = \int \frac{d^3 k}{(2\pi)^3} P_{\zeta'}(k) = \int_{k_{\rm min}}^{k_{\rm max}} \left[\frac{k^3}{2\pi^2} P_{\zeta'}(k)\right] d(\ln k)\, .
\end{equation}
We will thus take the square root of the variance per log-interval as our RMS estimator for the amplitude of the conformal time-derivative
\begin{equation}
\zeta_{\mathbf k,\rm inf}^{'\rm rms} \sim \sqrt{\frac{k^3}{2\pi^2} P_{\zeta'}(k)} 
= \sqrt{2 k^3 \sigma_p^2 \eta_i^2 \frac{H_{dS}^2}{8\pi^2 M_{pl}^2\epsilon}} = \sqrt{2 k^3 \sigma_p^2 \eta_i^2\ \Delta_\zeta^2(k)}\,,
\end{equation}
where we reorganized terms so that a factor of the curvature amplitude $\Delta_\zeta^2$ is apparent.
With this dictionary established, any model of decoherence from our landscape can be directly evaluated to find the initial velocity of the curvature perturbation at the onset of the radiation era.

For example, consider the momentum variance of a pure two-mode squeezed state in a inflationary background
\begin{equation}
    \sigma_{p,\rm pure}^2 = \sigma_{p,\rm vac}^2 = \frac{k}{2}\, .
\end{equation}
The RMS time-derivative of the curvature perturbation is then (it is convenient to write the result in terms of $\zeta^{'}_{\mathbf k,\rm inf}/k$, which has the same dimensions of $\zeta_{\mathbf k,\rm inf}$)
\begin{equation}
\label{eq:zetaPrimePure}
    \frac{\zeta^{'\rm rms}_{\mathbf k,\rm inf}}{k} = (k\eta_i)\, \Delta_{\zeta}(k)\,,
\end{equation}
which is small in the large wavelength $(k\eta_i)\rightarrow 0$ limit.
Alternatively, the variance for the thermal/classical state is $\sigma_{p,\rm th}^2 = k^2 \sigma_v^2$ with $\sigma_v^2 \approx e^{2r_k}/(2k) \approx 1/(2k^3\eta_i^2)$, which becomes
\begin{equation}
    \frac{\zeta^{'\rm rms}_{\mathbf k,\rm inf}}{k} = \Delta_{\zeta}(k) = \zeta_{\mathbf k,\rm inf}^{\rm rms}\,,
    \label{eq:ThermalRMS}
\end{equation}
so that the RMS scale of the amplitude and its derivative are set by the same (approximately) scale invariant spectrum.

\subsection{Radiation Era Matching Conditions}
\label{subsec:Matching}

To understand the physical consequences of these decohered initial conditions, we must evolve the state forward in time. 
We choose to evaluate the resulting observables specifically during the subsequent radiation-dominated era, rather than during inflation itself, because during inflation the exact physical nature of the environment causing the decoherence is unspecified. 
This environment may act as a continuous measurement or interaction that injects stochastic momentum noise into the system, so that we cannot guarantee that the standard Einstein constraint equations remain unmodified in this era. 
However, once inflation ends and the universe enters the radiation era, we assume these unknown environmental interactions have ceased. 
The radiation era provides a robust, well-understood background governed strictly by standard general relativity, making it the cleanest epoch in which to examine the implications of a large initial momentum variance on the decaying mode and the physical gravitational potential.

To propagate these initial conditions through the sharp transition from inflation to the radiation-dominated era, we must map them onto the Newtonian observables. Working in the longitudinal (Newtonian) gauge, we parameterize the scalar metric perturbations in the absence of anisotropic stress as
\begin{equation}
ds^2 = a(\eta)^2 \left[ -(1+2\Phi)d\eta^2 + (1-2\Phi)d\vec{x}^2 \right]\,,
\end{equation}
where $\Phi$ is the Bardeen gravitational potential.
Now transforming to Fourier space, the uniform-density curvature perturbation $\zeta_{\mathbf k}$ can be rewritten on large scales in terms of $\Phi_{\mathbf k}$ using the $00$ Einstein equation
\begin{equation}
\zeta_{\mathbf k} = \Phi_{\mathbf k} + \frac{1}{\epsilon} \left( \Phi_{\mathbf k} + \frac{\Phi'_{\mathbf k}}{\mathcal{H}} \right)\,,
\end{equation}
where $\mathcal{H} = a'/a$ is the conformal Hubble parameter and $\epsilon = 1 - \mathcal{H}'/\mathcal{H}^2$ is the equation of state parameter.
Multiplying by $\epsilon \mathcal{H}$ and rearranging terms yields a differential equation linking the two variables
\begin{equation}
\Phi'_{\mathbf k} = \epsilon \mathcal{H} \zeta_{\mathbf k} - (\epsilon+1) \mathcal{H} \Phi_{\mathbf k}\,.
\label{eq:PhiPrime}
\end{equation}
The perturbed Einstein equations (specifically the momentum constraint) connect the time-derivative of $\zeta_{\mathbf k}$ to the spatial Laplacian of the gravitational potential
\begin{equation}
\zeta'_{\mathbf k} = -\frac{c_s^2 k^2}{\epsilon \mathcal{H}} \Phi_{\mathbf k}\,,
\label{eq:ZetaPrime}
\end{equation}
where $c_s$ is the sound speed of the cosmological fluid.

These two first-order equations allow us to evolve the perturbations across the transition hypersurface between inflation ($\eta \to \eta_i^-$) and the radiation era ($\eta \to \eta_i^+$). 
The Deruelle-Mukhanov matching conditions \cite{Deruelle:1995kd} mandate that the induced metric and the extrinsic curvature of a transition hypersurface of constant energy density must be continuous. 
For scalar perturbations, this strict continuity imposes two fundamental matching conditions
\begin{align}
&\text{ Continuity of the Potential} \Longrightarrow \Phi_+ = \Phi_-  \\
&\text{ Continuity of the Curvature} \Longrightarrow \zeta_+ = \zeta_-
\end{align}
Notice that while the curvature perturbation itself must be continuous, there is no requirement for its derivative $\zeta'$ to be continuous. 
Indeed, by rearranging our momentum constraint \eqref{eq:ZetaPrime}, we can express the gravitational potential as $\Phi_{\mathbf k} = -\frac{\epsilon \mathcal{H}}{c_s^2 k^2} \zeta'_{\mathbf k}$, so that the
matching condition for the potential ($\Phi_+ = \Phi_-$) yields a matching condition for the derivative of the curvature perturbation across the transition
\begin{equation}
\left(\frac{\epsilon}{c_s^2} \zeta'\right)_+ = \left(\frac{\epsilon}{c_s^2} \zeta'\right)_-
\end{equation}
where we assumed the conformal Hubble paramater $\mathcal{H}$ is continuous across the transition.

We can now apply this to the specific transition from inflation to radiation. During inflation, we take the sound speed to be $c_s^2 = 1$ and $\epsilon = \epsilon_{\rm inf} \ll 1$. 
In the subsequent radiation era, the sound speed becomes $c_s^2 = 1/3$ and the equation of state mandates $\epsilon = 2$, so the matching conditions for the initial curvature perturbation $\zeta_{\mathbf k,r}(\eta_i)\equiv \zeta_i$ and its derivative $\zeta'_{\mathbf k,r}(\eta_i) \equiv \zeta_i'$ in the new radiation epoch are
\begin{equation}
\zeta_i = \zeta_{{\mathbf k},\rm inf} = \Delta_{\zeta}, \qquad \zeta'_i = \frac{\epsilon_{\rm inf}}{6} \zeta'_{{\mathbf k},\rm inf} = \frac{\epsilon_{\rm inf}}{6} \sqrt{2 k^3 \sigma_p^2 \eta_i^2\ \Delta_\zeta^2}\, .
\label{eq:MatchingFinal}
\end{equation}
Thus, the values of the curvature perturbation and its time derivative generated by decoherence at the end of inflation $(\zeta_{\mathbf k, \rm inf}^{\rm rms},\zeta^{'\rm rms}_{{\mathbf k},\rm inf})$ are transferred into the initial conditions of the curvature perturbation for the radiation era $(\zeta_i, \zeta'_i)$.

It is worth briefly commenting on the possibility of a transient matter-dominated phase (such as a period of perturbative reheating driven by a massive oscillating inflaton) preceding radiation domination. 
In a true dust-like, pressureless fluid, the sound speed vanishes ($c_s^2 \rightarrow 0$). 
Inspecting the momentum constraint governing the fluid \eqref{eq:ZetaPrime}, it becomes immediately apparent that any non-zero time derivative $\zeta'$ in a matter-dominated era formally sources a divergent gravitational potential $\Phi$. 
However, a transient early matter-dominated phase driven by a coherently oscillating scalar field should be able to absorb a non-zero $\zeta'$ gradient pressures of the scalar field perturbations.
To bypass these 
subtleties and evaluate a conservative, stable limit, we assume a direct transition to the radiation-dominated era ($c_s^2 = 1/3$), where acoustic pressure can safely absorb the kinematic kick.

\subsection{Decoherence Sources the Decaying Mode}

In order to see the effects of the initial conditions $(\zeta_i,\zeta'_i)$ generated by inflation and decoherence, we must match them onto the time-dependent solutions of the radiation-dominated universe.
Combining our constraint equations \eqref{eq:PhiPrime},\eqref{eq:ZetaPrime}, we obtain a single second order differential equation for the gravitational potential $\Phi_{\mathbf k}$
\begin{equation}
    \Phi''_{\mathbf k} + \left[(\epsilon+1)\mathcal{H} - \frac{\epsilon'}{\epsilon} - \frac{\mathcal{H}'}{\mathcal{H}}\right] \Phi'_{\mathbf k} + \left[c_s^2 k^2 - \frac{\epsilon'}{\epsilon}\mathcal{H}\right] \Phi_{\mathbf k} = 0\, .
\end{equation}
In the radiation era ($c_s^2 = 1/3, {\mathcal H} = 1/\eta$), this simplifies to
\begin{equation}
    \Phi''_{\mathbf k} + \frac{4}{\eta} \Phi'_{\mathbf k} + \frac{k^2}{3} \Phi_{\mathbf k} = 0\, .
\end{equation}
which has exact analytical solutions terms of spherical Bessel functions
\begin{equation}
    \Phi_{\mathbf k}(x) = A_{\mathbf k} \frac{j_1(x)}{x} + B_{\mathbf k} \frac{n_1(x)}{x}\,,
\end{equation}
with $x \equiv k\eta/\sqrt{3}$.
To map this to our boundary conditions, we evaluate these solutions at early times on superhorizon scales $x \ll 1$. Expanding the Bessel functions, the gravitational potential takes the asympototic form
\begin{equation}
    \Phi_{\mathbf k}(x) \approx \frac{A_{\mathbf k}}{3} + \frac{B_{\mathbf k}}{x^3}\, \qquad \text{for } x \ll 1\, .
    \label{eq:PhiRadiation}
\end{equation}
The coefficient $A_k$ controls the standard constant ``growing'' mode, while $B_k$ controls the singular ``decaying'' mode.
At later times, after the modes re-enter the horizon $x \gg 1$ the Bessel functions become oscillating solutions
\begin{equation}
    \Phi_{\mathbf k}(x) \approx -\left(A_{\mathbf k} \frac{\sin x}{x^2} + B_{\mathbf k} \frac{\cos x}{x^2}\right) \qquad \text{for }x \gg 1\, .
\end{equation}
The usual acoustic oscillations have a phase set by the $A_k$ growing modes; a decaying mode sourced at the same level as the growing mode $B_k\sim A_k$ would destroy the phase coherence of the CMB and be ruled out by observations.

Using the constraints \eqref{eq:PhiPrime},\eqref{eq:ZetaPrime} and with the initial conditions ($\zeta(\eta_i) = \zeta_i$ and $\zeta'(\eta_i) = \frac{\epsilon_{\rm inf}}{6} \zeta'_i$), we can solve for the growing and decaying mode coefficients
\begin{align}
    \label{eq:GrowingAmplitude}
    A_{\mathbf k} &= 2 \zeta_i \\
    \label{eq:DecayingAmplitude}
    B_{\mathbf k} &= -\frac{2}{9 \sqrt{3}} (k\eta_i)^3 \zeta_i - \frac{2}{\sqrt{3}} (k\eta_i)^2 \frac{\zeta_i'}{k}\,.
\end{align}
We can estimate the physical magnitude of these coefficients by substituting the stochastic RMS amplitude of the curvature perturbation $\zeta_i$ and its derivative $\zeta'_i$ through \eqref{eq:MatchingFinal}.
The result for the growing mode \eqref{eq:GrowingAmplitude} leads to the standard result for the relationship between the gravitational potential and the curvature perturbation on superhorizon scales at late times, $\Phi \approx (2/3) \zeta_i = (2/3) \Delta_\zeta$, for all decoherence models.
The decaying mode, however is directly sourced by the momentum variance from the decoherence model
\begin{equation}
    B_k = -\frac{2}{9\sqrt{3}} (k\eta_i)^3 \Delta_{\zeta} -  \frac{\epsilon_{\rm inf}}{3\sqrt{3}} (k\eta_i)^3 \Delta_{\zeta}\,\sqrt{\frac{\sigma_p^2}{\sigma_{p,\rm vac}^2}}\, .
    \label{eq:DecayingModeCoefficient}
\end{equation}
For example, for a pure two-mode squeezed state with a momentum variance equal to that of the vacuum $\sigma_p^2/\sigma_{p,\rm vac}^2 = 1$,
the resulting coefficient of the decaying mode is highly suppressed compared to the growing mode
\begin{equation}
    B_{\mathbf k}^{(\rm pure)} \approx -\frac{2}{9\sqrt{3}} (k\eta_i)^3\Delta_\zeta - \frac{1}{3\sqrt{3}} (k\eta_i)^3 \epsilon_{\rm inf} \Delta_\zeta \ll A_{\mathbf k}^{(\rm pure)}\, .
    \label{eq:DecayingModePure}
\end{equation}

The relationship between the canonical momentum variance $\sigma_p^2$ and the decaying mode \eqref{eq:DecayingModeCoefficient} exposes a tension between standard phenomenological practice and the rigorous requirements of quantum continuous-variable systems.
In the standard phenomenological approach to structure formation, the post-inflationary perturbations are treated as a classical stochastic ensemble \cite{Starobinsky:1980te, Kiefer:1998qe, Martin:2015qta}.
In this framework, as the Wigner ellipse becomes infinitely elongated due to inflationary squeezing, the momentum and amplitude become perfectly correlated. The Wigner function approaches a delta-function trajectory in phase space, $W(v,p) \propto \delta(p + v/\eta)$, seemingly allowing the decaying mode to be ignored without requiring physical environmental decoherence.
When initializing Boltzmann codes (e.g., CLASS \cite{CLASSI,CLASSII} or CAMB \cite{CAMB}) to calculate Cosmic Microwave Background observables, it is then standard practice to restrict this ensemble entirely to the growing mode, effectively setting the decaying mode to zero.
Indeed, we see that the decaying mode coefficient of the pure state \eqref{eq:DecayingModePure} is suppressed by powers of $(k\eta_i) \ll 1$, justifying such an approach.

\begin{figure}[t]
\centering\includegraphics[width=0.9\textwidth]{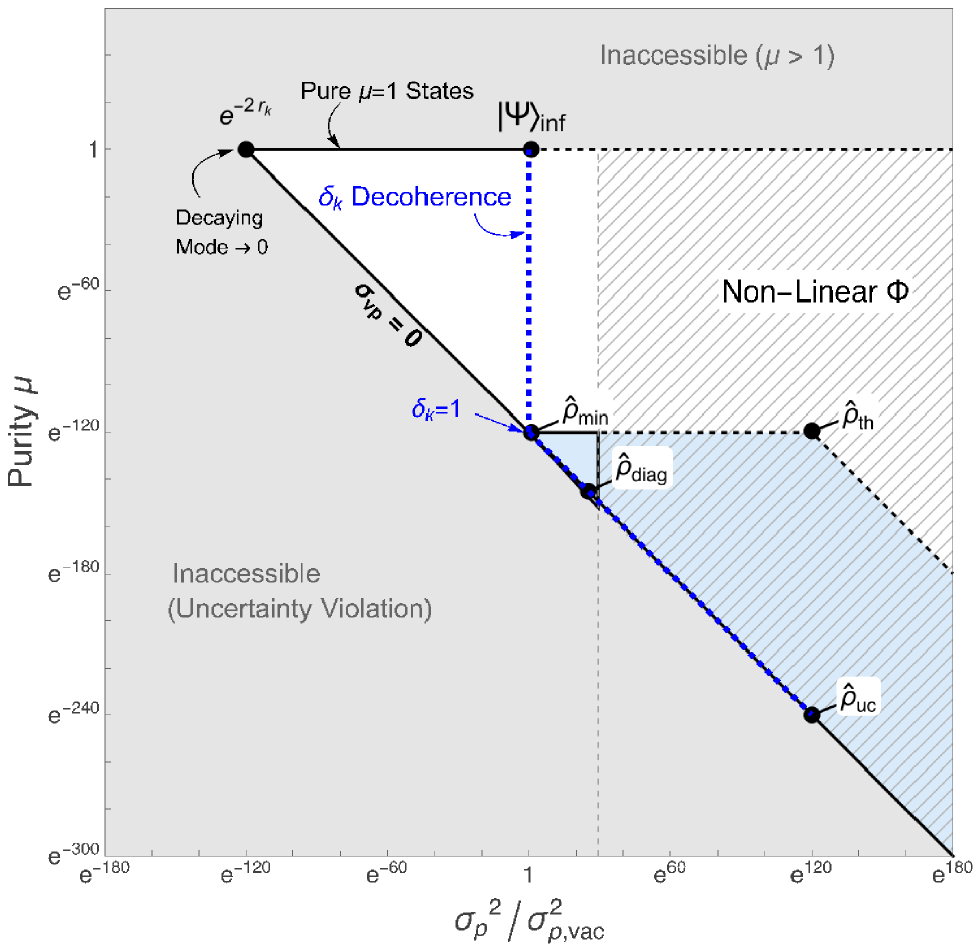}
\caption{The momentum variance created by decoherence must be below a certain threshold \eqref{eq:NonlinearBound} in order for the graviational potential $\Phi_{\mathbf k}(\eta_i)$ at the start of the radiation era to be linear $\Phi_{\mathbf k}(\eta_i) \ll 1$, shown by the vertical dashed line. Decoherence models with only classical correlations, such as the phase-averaged two-mode thermal state $\hat \rho_{\rm th}$ and the uncorrelated thermal state $\hat \rho_{\rm uc}$, are therefore ruled out as viable models of decoherence because they generate an exponentially large gravitational potential. There is only a narrow wedge of parameters that satisfy both this bound and can be described by a classical stochastic distribution (blue shaded region). The standard assumption of numerical CMB codes is to set the decaying mode of the gravitational potential to zero,  corresponding to a vanishingly small momentum variance in the upper-left corner of the landscape and a state that is quite far from the classical, blue-shaded region.}
\label{fig:LandscapeConstraints}
\end{figure}

However, the conclusion drawn from the quantum information perspective is starkly different. 
If this distribution is to be interpreted as a true classical statistical mixture, setting the decaying mode identically to zero implies a vanishing momentum variance, $\sigma_p^2 \to 0$. 
On our parameter landscape Figure \ref{fig:LandscapeConstraints}, this places the phenomenological universe in the extreme far-left region  -- a mathematically forbidden sector that violates the fundamental uncertainty bounds of the quantum vacuum. 
To consider a vanishingly small decaying mode while still maintaining a valid Gaussian state, the momentum variance is instead pushed to its absolute minimum, $\sigma_p^2/\sigma_{p,\rm vac}^2 \rightarrow e^{-2r_k}$, in the extreme upper-left corner of the allowed landscape, Figure \ref{fig:LandscapeConstraints}.
Notice that this location is as far as theoretically possible from the ``classical" states (the shaded region) that possess a regular, non-negative $P$-function.
Interestingly, this implies that any model of ``decoherence" that strongly suppresses the decaying mode in this way retains essential quantum signatures.

To locate a physically consistent, decohered mixed state of cosmological perturbations on our landscape possessing a regular, non-negative $P$-function, the state must reside in the shaded blue region of Figure \ref{fig:LandscapeConstraints}. 
Rather than suppressing the decaying mode, this decoherence requires physical environmental decoherence to inject stochastic noise into the momentum, and thereby the decaying mode, suppressing the purity to $\mu \lesssim e^{-2r_k}$ while simultaneously driving the momentum variance above the vacuum limit ($\sigma_p^2 > \sigma_{p,\text{vac}}^2$).

Let us consider the effects that an enhanced momentum variance can have on the CMB and gravitational dynamics through \eqref{eq:DecayingModeCoefficient}.
One consequence of models of decoherence that lead to a large momentum variance may be
the loss of the phase coherence of the CMB \cite{Dodelson:2003ip, Campo:2003pa, Campo:2004sz, Kiefer:2008ku, Micheli:2022tld}.
Let's examine this in more detail by considering the size of the decaying mode coefficient for a thermal density matrix (e.g. \eqref{eq:2ModeThermal} or \eqref{eq:ClassicalDensity}), the decoherence model considered here with the largest momentum variance.
Using \eqref{eq:ThermalRMS} and matching across the inflation to radiation transition, we have $\zeta_i'/k = (\epsilon_{\rm inf}/6)$ and $ \zeta_{\mathbf k, \rm inf}^{'\rm rms}/k = (\epsilon_{\rm inf}/6) \zeta_i$.
Using these in \eqref{eq:GrowingAmplitude},\eqref{eq:DecayingAmplitude}, we see that the coefficient of the decaying mode is still strongly suppressed compared to the growing mode
\begin{equation}
    B_{k}^{\rm(th)} \approx -\epsilon_{\rm inf} (k\eta_i)^2 \zeta_i = \frac{3}{2} \epsilon_{\rm inf} (k\eta_i)^2 A_{k}^{\rm (th)}\, .
    \label{eq:DecayingModeCoefficientThermal}
\end{equation}
Because of the strong suppression from $\epsilon_{\rm inf} (k\eta_i)^2 \ll 1$, there should be no loss of phase coherence for the acoustic oscillations when the modes enter the horizon.
Since the thermal/classical density matrix has the largest momentum variance $\sigma_p^2$ of the decoherence models we considered, we see that none of the decoherence models will induce a loss of phase coherence in the CMB.
This is consistent with other, independent, studies of the impact of the decaying mode on the phase coherence of the CMB \cite{Amendola:2004rt, Kodwani:2019ynt, dePutter:2019xxv}.

However, while the decaying mode does not spoil the phase coherence of the CMB, it does place severe theoretical constraints on the viability of decoherence models through the gravitational dynamics at the reheating matching surface $\eta_i$ \cite{Campo:2005sy}.
As noted before, the Einstein equations in the radiation era give us at the matching surface the exact relation
\begin{equation}
    \Phi_{\mathbf k}(\eta_i) = -\frac{6}{k\eta_i} \frac{\zeta_i'}{k}\, .
    \label{eq:PhiZetaPrime}
\end{equation}
As before, we estimate the physical magnitude of the potential by substituting the stochastic RMS amplitude $\zeta_i'$ from \eqref{eq:MatchingFinal}.
For the pure two-mode squeezed state the derivative of the curvature perturbation at the end of inflation is vanishingly small in the superhorizon limit \eqref{eq:zetaPrimePure}, which combined with the matching condition \eqref{eq:MatchingFinal} gives the gravitational potential
\begin{equation}
    \Phi_{\mathbf k}(\eta_i) = - \epsilon_{\rm inf} \Delta_{\zeta} \qquad \text{(Pure Two-Mode)}\, .
\end{equation}
Recall that $\Delta_{\zeta} \sim 10^{-5}$, and $\epsilon_{\rm inf} \sim 10^{-2}$, so the gravitational potential at the start of the radiation era is suppressed $\Phi_{\mathbf k}(\eta_i) \sim 10^{-7}$.
Note that this represents the value of the gravitational potential precisely at the reheating hypersurface matching the inflationary and radiation eras, in which the contribution of the decaying mode partially cancels the growing mode.
At later times in the radiation era the decaying mode will decay away, and the superhorizon gravitational potential will be dominated by the (constant) growing mode $\Phi_{\mathbf k}(\eta>\eta_i) \approx (2/3) \Delta_{\zeta}$.

Consider again the thermal density matrix (\eqref{eq:2ModeThermal} or \eqref{eq:ClassicalDensity}) in which 
\begin{equation}
\frac{\zeta_i'}{k} = \frac{\epsilon_{\rm inf}}{6} \frac{\zeta_{\mathbf k, \rm inf}^{'\rm rms}}{k} = \frac{\epsilon_{\rm inf}}{6} \zeta_{\mathbf k,\rm inf}^{\rm rms} = \frac{\epsilon_{\rm inf}}{6} \Delta_{\zeta} \quad \text{(Thermal State)}\,.
\end{equation}
For superhorizon modes, $k\eta_i$ is approximately equal to the squeezing of the state $(k|\eta_i|) \approx e^{-r_k}$ with $r_k \approx N_k$ equal to the number of e-folds since horizon exit, so with $N_k \sim 60$ and $\epsilon_{\rm inf} \sim 10^{-2}$, we find that the gravitational potential \eqref{eq:PhiZetaPrime} is highly non-linear
\begin{equation}
    \Phi_{\mathbf k}(\eta_i) = -\frac{\epsilon_{\rm inf} \Delta_\zeta}{k\eta_i} \sim 10^{19} \gg 1\qquad \text{(Thermal State)}\, .
\end{equation}
Thermal state decoherence generates too large of a gravitational potential at the reheating surface, violating the linear analysis and perturbation theory at reheating.

More generally, using \eqref{eq:MatchingFinal}, we can write the gravitational potential in terms of the momentum variance for any model of decoherence
\begin{equation}
    \Phi_{\mathbf k}(\eta_i) = \epsilon_{\rm inf} \sqrt{\frac{2\sigma_p^2}{k}}\, \Delta_{\zeta} = \epsilon_{\rm inf} \sqrt{\frac{\sigma_p^2}{\sigma_{p,\rm vac}^2}} \Delta_\zeta\, .
    \label{eq:PhiSigmaP}
\end{equation}
For linear perturbation theory to remain valid, the initial gravitational potential must remain well below order unity ($\Phi_{\rm rms} \ll 1$). This requirement imposes an upper bound on the momentum variance generated by any decoherence mechanism
\begin{equation}
    \frac{\sigma_p^2}{\sigma_{p,\rm vac}^2} \ll (\epsilon_{\rm inf}^2\Delta_\zeta^2)^{-1} \sim 5\times 10^{12} \approx e^{29.2}\, . 
    \label{eq:NonlinearBound}
\end{equation}
A stronger requirement was advocated by Campo and Parentani \cite{Campo:2005sy}, who demanded that the power of the decaying mode remain below that of the growing mode at the onset of the radiation era, arguing that the Bardeen potential should not exceed its standard in-vacuum growing-mode value.
Imposing the analogous condition within our framework, 
$\Phi_{\rm dec}(\eta_i) \lesssim \Phi_{\rm grow}(\eta_i) = \frac{2}{3} \Delta_{\zeta}$, 
and using \eqref{eq:PhiSigmaP}, gives
\begin{equation}
\frac{B_{\mathbf k}}{(k\eta_i)^3} \lesssim A_{\mathbf k}\,, \qquad \Rightarrow \qquad \frac{\sigma_p^2}{\sigma_{p,\rm vac}^2} \lesssim \epsilon_{\rm inf}^{-2} \sim 10^4 \sim e^{9.2}\, ,
 \label{eq:CampoParentaniBound}
\end{equation}
which is significantly more stringent than our linearity bound \eqref{eq:NonlinearBound} by a factor of $\Delta_\zeta^{-2}$.
We will primarily focus on the weaker non-linearity bound \eqref{eq:NonlinearBound} as the physically necessary one, while noting that the stronger condition \eqref{eq:CampoParentaniBound} may lead to more stringent constraints on decoherence models.

\begin{figure}[t]
\centering\includegraphics[width=0.9\textwidth]{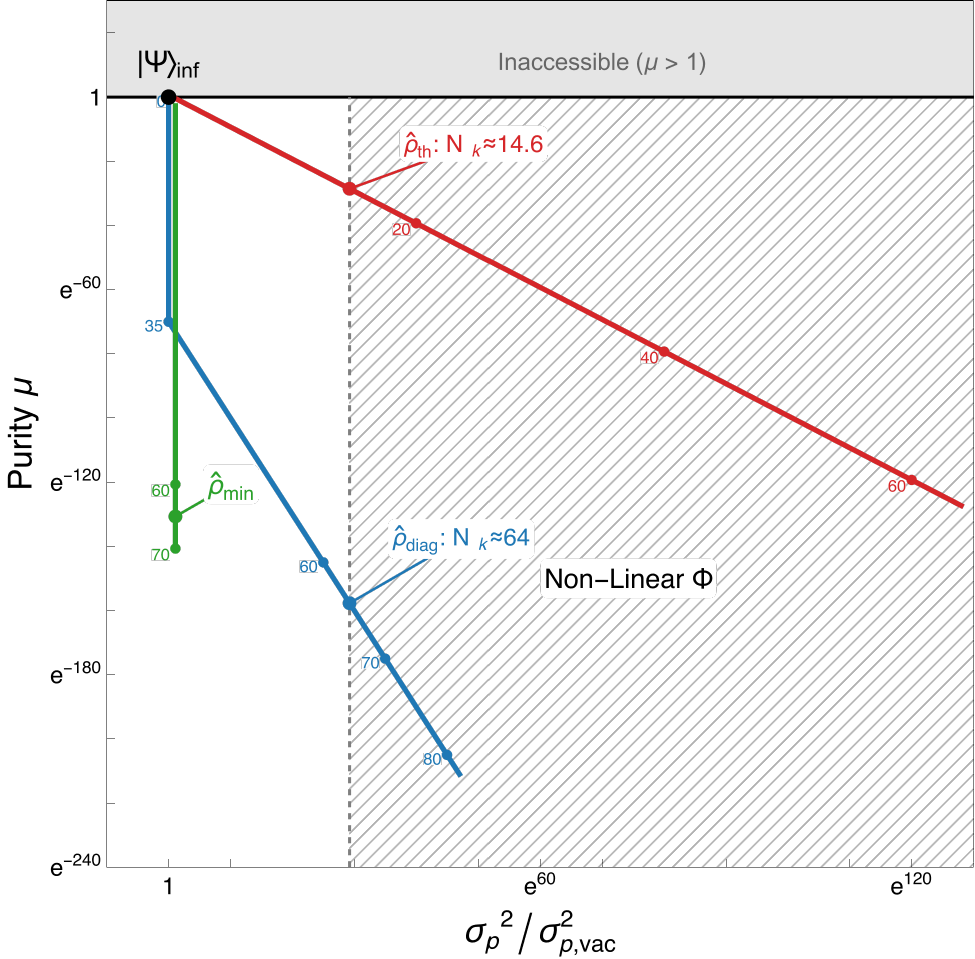}
\caption{The momentum variance for the phase-averaged thermal state $\hat \rho_{\rm th}$ and the diagonal decoherence $\hat \rho_{\rm diag}$ depends on the number of e-folds a decohered mode is outside of the horizon. When the number of e-folds is large enough, the momentum variance can exceed the bound \eqref{eq:NonlinearBound} beyond which the gravitational potential becomes non-linear in the radiation era. For the thermal state, the bound is exceeded after only $N_k \approx 14.6$ e-folds, while for the diagonal decoherence model the bound is exceeded after $N_k \approx 64$ e-folds. The minimal decoherence model $\hat \rho_{\rm min}$ is also shown here for comparison; the momentum variance is independent of the e-folds for this model, while the purity decreases with increasing e-folds.}
\label{fig:Nk_Landscape}
\end{figure}

Let's compare the model of decoherence that is diagonal in amplitude basis \eqref{eq:DiagAmplitudeDensity} with momentum variance \eqref{eq:DiagonalMomentumVariance} 
\begin{equation}
     \frac{\sigma_{p,\text{diag}}^2}{\sigma_{p,\text{vac}}^2} = \frac{1}{288}(\epsilon+\eta)^2\Delta_{\zeta}^2\ e^{N_k} \approx e^{r_k-35}\, ,
    \label{eq:DiagonalMomentumVariance2}
\end{equation}
to the bound \eqref{eq:NonlinearBound}, where we used that
the squeezing parameter is approximately equal to the number of e-folds a mode has been outside the horizon ($r_k \approx N_k$).  
This model satisfies the bound \eqref{eq:NonlinearBound} as long as $r_k \approx N_k \lesssim 64$, and this result is only logarithmically sensitive to changes in the slow roll parameter $\epsilon_{\rm inf}$.
This suggests that for the diagonal decoherence model to describe the decoherence of cosmological perturbations, inflation must end after a maximum of $64$ e-folds or risk sourcing a non-linear gravitational potential at the start of the radiation era.
The minimal decoherence model \eqref{eq:MinimalTwoModeCoherentDensity} sits far below the bound \eqref{eq:NonlinearBound} at $\sigma_{p,\rm min}^2/\sigma_{p,\rm vac}^2 = 3$, injecting only a vacuum-level correction into the decaying mode, while the $\delta_k$-decoherence models with $\sigma_{p,\delta}^2/\sigma_{p,\rm vac}^2 = 1+\delta_k$ sit below the bound for $\delta_k \lesssim 10^{12}$.
Thus, the requirement that the gravitational potential be small and in the linear regime puts strong constraints on the allowed space of decoherence models.

Superimposing the bound \eqref{eq:NonlinearBound} on our landscape of allowed decoherence models, Figure \ref{fig:LandscapeConstraints}, 
we see that any model to the right of this line injects so much energy into the decaying mode that the gravitational potential becomes non-linear.
This constraint rules out a significant fraction of the available parameter space. 
In particular, it excludes mixed states with classical-only correlations, such as the phase averaged two-mode thermal state $\hat \rho_{\rm th}$ \cite{Brandenberger:1992sr,Brandenberger:1992jh,Prokopec:1992ia,Gasperini:1992xv,Gasperini:1993mq,Brahma:2020zpk} and the uncorrelated thermal state $\hat \rho_{\rm uc}$.
Combining this with the requirement that the state be P-representable, leaves only a narrow wedge (narrow, at least, in log variables) in which a decoherence model can avoid
generating too large a gravitational potential while remaining classically representable.
This wedge contains the minimal decoherence model $\hat \rho_{\rm min}$ of \eqref{eq:MinimalTwoModeCoherentDensity}, the $\delta_k$-decoherence model \eqref{eq:deltaCoherenceDefinition} for $\delta_k \lesssim 10^{12}$, and the model of decoherence in which the density matrix is diagonal in amplitude-basis \eqref{eq:DiagAmplitudeDensity} for modes that have been superhorizon for up to $\sim 64$ e-folds.

While we have generally evaluated the decoherence models of Section \ref{sec:decoherence} for modes that have been outside the horizon for a nominal $r_k \sim N_k \sim 60$ e-folds, it is instructive to let the amount of squeezing (and hence the number of e-folds) experienced by a given $k$-mode vary. 
Each model then traces out a trajectory across the landscape as a function of $N_k$.
For example, the diagonal decoherence depends on the number of e-folds as
\begin{equation}
    \left(\sigma_{p,\rm diag}^2/\sigma_{p,\rm vac}^2,\ \mu_{\rm diag}\right) = \left(1+e^{N_k-35},e^{-(3N_k - 35)}\right)\, ,
\end{equation}
so that the non-linearity threshold \eqref{eq:NonlinearBound} is crossed at $N_k \lesssim 35 + 29.2 \approx 64$.
The corresponding expression for phase-averaged thermal states is
\begin{equation}
    \left(\sigma_{p,\rm th}^2/\sigma_{p,\rm vac}^2,\ \mu_{\rm th}\right) = \left(e^{2N_k},2e^{-2N_k}\right)\, ,
\end{equation}
so that the non-linearity threshold \eqref{eq:NonlinearBound} is crossed at a much lower number of e-folds $N_k \lesssim 29.2/2 = 14.6$, far from the $60$ e-folds typically needed for inflation.
In contrast, the momentum variance for the minimal decoherence model is constant
\begin{equation}
    \left(\sigma_{p,\rm min}^2/\sigma_{p,\rm vac}^2,\ \mu_{\rm min}\right) = \left(3,e^{-2N_k}/2\right)\, ,
\end{equation}
so that this model never crosses the non-linearity threshold (a similar expression holds for the $\delta_k$-decoherence model, with $\sigma_{p,\delta}^2/\sigma_{p,\rm vac}^2 = 1+\delta_k$).
These trajectories show how each model scans across the landscape with the number of e-folds; see Figure \ref{fig:Nk_Landscape}.

A complementary and physically revealing way to visualize the violation of the the non-linearity bound for the thermal and diagonal decoherence models
is to examine the  real-space variance of the gravitational potential
\begin{equation}
\langle \Phi(\mathbf x)^2\rangle = \int \frac{d^3k}{(2\pi)^3} P_{\Phi}(k) = \int_{k_{\rm min}}^{k_{\rm max}} \left[\frac{k^3}{2\pi^2} P_{\Phi}(k)\right] d(\ln k)\, .
\label{eq:PhiRealCorrelation1}
\end{equation}
Using the classical boundary condition \eqref{eq:PhiZetaPrime} relating the potential to the curvature derivative, the power spectrum of the gravitational potential scales as $P_{\Phi}(k) \propto k^{-4} P_{\zeta'}(k)$.
Substituting our the two-point statistics of $\zeta_i'$ \eqref{eq:zetaPrimePowerSpectrum2}
into \eqref{eq:PhiRealCorrelation1}, the real-space variance becomes an integral over the normalized momentum variance
\begin{equation}
    \langle \Phi(\mathbf x)^2\rangle = \int_{k_{\rm min}}^{k_{\rm max}} \left[\frac{k^3}{2\pi^2} \frac{\epsilon_{\rm inf}^2}{(k\eta_i)^2} \frac{P_{\zeta'}(k)}{k^2}\right]d(\ln k) = \int_{k_{\rm min}}^{k_{\rm max}} \left[\Delta_\zeta^2 \epsilon_{\rm inf}^2 \frac{\sigma_p^2}{\sigma_{p,\rm vac}^2}\right] d(\ln k)\, .
    \label{eq:PotentialRealVariance}
\end{equation}
The bound \eqref{eq:NonlinearBound} thus also ensures the real-space variance is small and in the linear regime.

Expressed in this form, the variance \eqref{eq:PotentialRealVariance} exposes a potential difficulty regarding the infrared (IR) regularity of decoherence models. 
Because the dimensionless curvature power spectrum $\Delta_\zeta^2(k)$ is approximately scale-invariant across cosmological scales, the convergence of this integral at large scales depends entirely on the $k$-dependence of the momentum variance. 
To avoid an IR singularity as we integrate over increasingly large superhorizon scales ($k \to 0$), the ratio $\sigma_p^2/\sigma_{p,\rm vac}^2$ must either remain strictly bounded or vanish in the infrared limit.
Different models on our landscape exhibit different asymptotic behaviors in this limit. 
Because the the momentum variance of the pure two-mode squeezed state is equal to the vacuum value, the integrand of \eqref{eq:PotentialRealVariance} is scale-invariant and safe from IR divergences.
Similarly, the minimal decoherence and generalized $\delta_k$-decoherence models inject only a scale-invariant, vacuum-level variance ($\sigma_p^2/\sigma_{p,\rm vac}^2 \sim \mathcal{O}(1)$), resulting in a scale-invariant integrand. 
However, models driven by continuous environmental interactions have potential challenges in the infrared. 
The momentum variance of the phase-averaged thermal state scales as $e^{2r_k} \propto k^{-2}$, while the dynamically rigorous amplitude-diagonal model scales as $e^{r_k} \propto k^{-1}$. 
For these models, the integrand of the variance \eqref{eq:PotentialRealVariance} diverges as $k \to 0$.
Thus, models that lie deep within the classical stochastic regime of our parameter space are potentially disfavored by infrared divergences in the momentum variance.
It would be interesting to study these models further to see if this scale dependence truly poses a challenge by generating non-linear gravitational potentials.

In addition to violating the linearity condition, a large decaying mode at the start of the radiation era could affect the local expansion of a patch of the universe, potentially shutting down expansion altogether.
In Newtonian gauge, the local expansion rate is given by
\begin{equation}
    H_{\rm loc} = \frac{1}{\sqrt{1+2\Phi}} \left[\frac{\dot a}{a} - \frac{\dot \Phi}{1-2\Phi}\right] = \frac{1}{\sqrt{2\Phi}} \frac{1}{a} \left[\frac{a'}{a} - \frac{\Phi'}{1-2\Phi}\right]\,,
\end{equation}
where we transformed from derivatives with respect to cosmic time (denoted by a dot) to derivatives with respect to conformal time (denoted by a prime).
Evaluating the gravitational potential and its derivative at the start of the radiation era 
\begin{equation}
    \Phi(\eta_i) = -\frac{6}{k\eta_i} \frac{\zeta_i'}{k}, \qquad \Phi'(\eta_i) \approx \frac{18 k}{(k\eta_i)^2} \frac{\zeta_i'}{k}\, ,
\end{equation}
the local expansion rate becomes
\begin{equation}
    H_{\rm loc} = \frac{1}{\sqrt{1+2\Phi}} \frac{1}{a\eta_i} \left[1-\frac{\frac{3}{k\eta_i} \frac{\zeta_i'}{k}}{1+\frac{2}{k\eta_i} \frac{\zeta_i'}{k}}\right]\, .
\end{equation}
For $\frac{1}{k\eta_i} \frac{\zeta_i'}{k} \gg 1$ (which from \eqref{eq:PhiZetaPrime} is the same as $\Phi_{\mathbf k} \gg 1$), the local expansion rate is negative.
Thus, a sufficiently large excitation of the decaying mode from decoherence not only violates linearity, but may shut down the local expansion, though this deserves further study.

\section{Summary and Conclusion}
\label{sec:conclusion}

Cosmic inflation predicts that primordial perturbations are described by a Gaussian density matrix $\hat{\rho}$ representing adiabatic fluctuations with a nearly scale-invariant amplitude. 
Subject to these observational constraints, we demonstrated that any generic Gaussian mixed state can be fully specified by two dimensionless parameters: the purity $\mu = \text{Tr}(\hat \rho^2)$ and the momentum variance relative to the vacuum, $\sigma_p^2/\sigma_{p,\rm vac}^2$.
Mapping the allowable parameter space reveals a comprehensive landscape of potential mixed states consistent with observations. 
This geometric framework unifies the literature on cosmological decoherence, allowing us to strictly locate distinct pointer-basis models as specific loci and trajectories within the overall landscape.

While many quantum-informational diagnostics of ``classicality" depend heavily on basis choice or arbitrary subsystem partitions, we imposed a strict, operational threshold: for the system to be mathematically representable by a regular, positive-definite classical stochastic probability distribution (the Glauber-Sudarshan $P$-function) whose variance exceeds that of the quantum vacuum.
This imposes a profound physical requirement. 
To cross the classicality threshold, the environment must actively inject momentum into the system above the quantum vacuum limit, thereby enhancing, rather than suppressing, the curvature decaying mode. 
This contrasts sharply with standard cosmological practice in which
phenomenological models and numerical Bolzmann solvers routinely assume that a classical stochastic ensemble emerges precisely when the decaying mode, and its associated momentum variance, is neglected \cite{Polarski:1995jg,Kiefer:2008ku}.
However, from a quantum information perspective discarding the decaying mode does not correspond to a mathematically realizable mixed state, while assuming it to be exponentially small drives the state into a highly quantum corner of the landscape.
Decoherence does not occur when the decaying mode is neglected; rather, decoherence requires the environment to inject energy into the momentum of the system, exciting the decaying mode.
For the system to reach the exponentially small purity $\mu \lesssim e^{-120}$ required for the system to be ``classical," the 
active interaction with the environment must generate a large von Neumann entropy per mode of $S_{\mathbf k} \sim \ln(1/\mu) \sim 120$, which may have implications for the early Universe.

We showed that the enhanced momentum variance due to decoherence translates directly into a macroscopic initial velocity for the curvature perturbation ($\zeta'$) at the transition into the radiation era. 
Through the Einstein equations, this injected momentum backreacts severely on the background geometry, acting as a source for the decaying mode of the Newtonian gravitational potential $\Phi$. 
While the kinematic suppression of this decaying mode ensures that it safely vanishes before recombination -- thus protecting the temporal coherence of the CMB acoustic peaks \cite{Dodelson:2003ip} -- its initial amplitude at the reheating surface places stringent theoretical constraints on the viability of decoherence models.

By requiring that $\Phi$ remain safely within the bounds of linear perturbation theory
($\Phi_{\rm rms} \ll 1$), we established an absolute upper bound on the allowable momentum variance.
Because the momentum variance of each decoherence model grows with the squeezing -- and hence with the number of e-folds $N_k$ a mode has spent outside the horizon -- this bound translates directly into a maximum number of e-folds each model can tolerate before the gravitational potential becomes non-linear. 
Highly decohered thermal states fare the worst: the phase-averaged thermal state violates the
bound already at $N_k \approx 14.6$ e-folds, so it is definitively ruled out for any mode of
cosmological interest.
Dynamically rigorous models that diagonalize the density matrix in the field amplitude basis do
somewhat better, but suffer from potentially catastrophic infrared divergences. 
Keeping the amplitude-diagonal model within the linear regime restricts long wavelength modes to having
been superhorizon for at most $\sim 64$ e-folds, capping the total duration of inflation at roughly the same value, a theoretical bound surprisingly close to the minimum $55$--$60$ e-folds required to solve the horizon problem.
Conversely, the pure inflationary quantum state and the ``minimal decoherence'' model (diagonal in the coherent-state basis) inject only vacuum-level momentum independent of $N_k$, and so remain completely safe from gravitational non-linearities and IR divergences no matter how long the corresponding mode is superhorizon.

The macroscopic excitation of the decaying mode by environmental decoherence potentially opens several new observational windows. While the linear potential rapidly decays, a transient epoch with $\Phi_{\rm rms} \sim 0.1$ approaches the critical threshold for local collapse, suggesting that severe decoherence could trigger a massive overproduction of Primordial Black Holes (PBHs). 
Furthermore, the large initial velocity $\zeta'$ may source appreciable tensor perturbations at second order. 
Evaluating whether specific models on the decoherence landscape overproduce a stochastic background of secondary gravitational waves remains a compelling target for future constraints on the quantum-to-classical transition of the early universe.

\acknowledgments
We would like to thank Amaury Micheli and Vincent Vennin for helpful conversations. 
BU would like to thank Natalie Elskamp for comments on an earlier version of this manuscript.
SSH is supported in part by the National Institute for Theoretical and Computational Sciences
of South Africa (NITheCS). BU would like to thank NITheCS for funding for travel to the University of Cape Town where part of this work was completed.

\appendix

\section{de Sitter and the Vacuum Momentum Variance}
\label{app:MomentumVariance}

In Section \ref{sec:review_pure} we saw that the momentum variance of the pure state on a de Sitter background was exactly equal to its vacuum value. 
To understand this effect, we start with the operator $\hat{v}_{\mathbf{k}}$ which obeys the well-known Mukhanov-Sasaki equation
\begin{equation}
    \hat{v}_{\mathbf k}'' + \left( k^2 - \frac{z''}{z} \right) \hat{v}_{\mathbf k} = 0\, .
    \label{eq:MukhanovSasakiEq}
\end{equation}
Let us introduce the operators
\begin{align}
    \hat A & = \frac{\partial}{\partial \eta} - \frac{z'}{z}\,; \\
    \hat A^\dagger & = -\frac{\partial}{\partial \eta} - \frac{z'}{z}\, .
\end{align}
The equation of motion \eqref{eq:MukhanovSasakiEq} can then be written as
\begin{equation}
    \hat A^\dagger \hat A\, \hat v_{\mathbf k} = k^2 \hat v_{\mathbf k}\, .
    \label{eq:MukhanovSasakiEqAA}
\end{equation}
Now we notice that the conjugate momentum \eqref{eq:MukhanovMomentum} is precisely $\hat p_{\mathbf k} = \hat A \hat v_{\mathbf k}$, so that by multiplying \eqref{eq:MukhanovSasakiEqAA} by $\hat A$ on the left we get
\begin{equation}
    \hat A \hat A^\dagger \left(\hat A \hat v_{\mathbf k}\right) = k^2 \left(\hat A \hat v_{\mathbf k}\right) \quad \Rightarrow \quad \hat A \hat A^\dagger \hat p_{\mathbf k} = k^2 \hat p_{\mathbf k}\, .
    \label{eq:MukhanovSasakiEqMomentum}
\end{equation}
The corresponding operator $\hat A \hat A^\dagger$ acting on $\hat p_{\mathbf k}$ becomes
\begin{equation}
    \hat A \hat A^\dagger = -\frac{\partial^2}{\partial \eta^2} + \left(W^2 - \frac{dW}{d\eta}\right)\, .
\end{equation}
For de Sitter we have $W = z'/z = -1/\eta$ so $W^2 - W' = 0$ and the equation of motion for the momentum is that of a free harmonic oscillator
\begin{equation}
    -\hat p_{\mathbf k}'' = k^2 \hat p_{\mathbf k}\, .
\end{equation}
As a result, by choosing the initial state in the Bunch-Davies vacuum, the momentum variance stays fixed at the vacuum value, $\sigma_p^2 = k/2$.

If we perturb slightly away from exact de Sitter space, this symmetry is broken by the slow roll parameters.
In particular, we can parameterize the deviation from de Sitter as
\begin{equation}
    \frac{z'}{z} \approx -\frac{1+\delta}{\eta}\,,
\end{equation}
where $\delta \sim {\mathcal O}(\epsilon,\eta)$ is set by the slow roll parameters of the quasi de Sitter background, with $|\delta| \ll 1$.
The equation of motion for the momentum \eqref{eq:MukhanovSasakiEqMomentum} is no longer free
\begin{equation}
    \hat p_{\mathbf k}'' + \left(k^2 - \frac{\delta}{\eta^2}\right) \hat p_{\mathbf k} = 0\, .
    \label{eq:MukhanovSaskaiSlowRoll}
\end{equation}
As a result, the momentum now feels a small, time-dependent effective mass proportional to the slow-roll parameters. 
Because of this, the momentum variance will begin to drift slightly from the vacuum value on superhorizon scales $k|\eta| \ll 1$.
In particular, the solution to \eqref{eq:MukhanovSaskaiSlowRoll} takes the form of a Hankel function
\begin{equation}
    \hat p_{\mathbf k} \sim \sqrt{k|\eta|}\, H_\nu^{(1)}(k|\eta|)\, ,
\end{equation}
where $\nu \approx 1/2 + \delta$.
In the superhorizon limit, the Hankel function scales as $(k|\eta|)^{-\nu}$, so the momentum variance $\sigma_p^2 = \langle \hat p_{\mathbf k}^2\rangle$ for quasi de Sitter (slow roll inflation) becomes
\begin{equation}
    \sigma_{p,\rm SR}^2 \approx \frac{k}{2} (k|\eta|)^{-2\delta} \approx \frac{k}{2} e^{2\delta \Delta N}\,,
\end{equation}
where $\Delta N$ is the number of efolds since horizon crossing.
For $\epsilon \sim 0.01$ and $\Delta N \sim 60$ efolds of inflation, the momentum variance is only a factor of a few larger than the vacuum value $\sigma_{p,\rm SR}^2 \sim 3.3 (k/2)$.

\bibliographystyle{JHEP}

\bibliography{refs}

\end{document}